\documentclass[5p,twocolumn,times]{elsarticle}

\usepackage{lineno,hyperref}
\usepackage{xcolor}
\usepackage{graphicx}
\modulolinenumbers[5]

\journal{Journal of \LaTeX\ Templates}

\biboptions{authoryear}

\def\ergscm{erg~s$^{-1}$~cm$^{-2}$}

\def\lum{erg s$^{-1}$}

\newcommand\aap{A\&A}                
\newcommand\ao{Appl. Opt.}           
\newcommand\apjl{ApJ}                
\newcommand\apjs{ApJS}               

\newcommand{\be}{\begin{equation}}
\newcommand{\ee}{\end{equation}}

\def\integral{\textit{INTEGRAL}}
\def\jemx{JEM-X}
\def\ibis{IBIS/ISGRI}
\def\spi{SPI}
\def\nh{NH}

\def\nustar{\textit{NuSTAR}}

\begin{document}

\begin{frontmatter}

\title{15 years of Galactic surveys and hard X-ray Background measurements}


\author[iki]{Roman A. Krivonos\corref{mycorrespondingauthor}}
\address[iki]{Space Research Institute (IKI), Profsoyuznaya 84/32, Moscow 117997, Russia}
\address[southampton]{School of Physics and Astronomy, University of Southampton, SO17 1BJ, UK}
\address[mpa]{Max-Planck-Institut f\"{u}r Astrophysik (MPA), Karl-Schwarzschild-Strasse 1, Garching 85741, Germany}
\address[berkeley]{Space Sciences Laboratory, 7 Gauss Way, University of California, Berkeley, CA 94720-7450, USA}
\address[inaf]{INAF - Istituto di Astrofisica e Planetologia Spaziali, Via Fosso del Cavaliere 100, Roma, I-00133, Italy}
\address[mesri]{Minist\`ere de l'Enseignement sup\'erieur, de la Recherche et de l'Innovation, 1 rue Descartes, 75005 Paris, France}
\address[esac]{European Space Astronomy Centre - ESA/ESAC, Villanueva de la Ca\~nada, Madrid, Spain}
\address[georgia]{Georgia College \& State University, CBX 82, Milledgeville, GA 31061, USA}
\address[aim]{AIM, CEA, CNRS, Universit\'{e} Paris-Saclay, Universit\'{e} de Paris, F-91191 Gif-sur-Yvette, France}
\address[apc]{Universit\'{e} de Paris, CNRS, Astroparticule et Cosmologie, F-75006 Paris, France}
\address[esa]{ESA-ESTEC, Research and Scientific Support Dept., Keplerlaan 1, 2201 AZ, Noordwijk, The Netherlands}
\address[oas]{INAF - Osservatorio di Astrofisica e Scienza dello Spazio, via Piero Gobetti 101, I-40129, Bologna, Italy}
\address[unab]{Departamento de Ciencias F\'isicas, Universidad Andr\'es Bello, Fern\'andez Concha 700, Las Condes, Santiago, Chile}
\address[utu]{Department of Physics and Astronomy, FI-20014 University of Turku,  Finland}

\cortext[mycorrespondingauthor]{Corresponding author}
\ead{krivonos@cosmos.ru}

\author[southampton]{Antony J. Bird}
\author[iki,mpa]{Eugene M. Churazov}
\author[berkeley]{John A. Tomsick}

\author[inaf]{Angela Bazzano}
\author[mesri]{Volker Beckmann}
\author[esac]{Guillaume B\'elanger}
\author[georgia]{Arash Bodaghee}
\author[aim,apc]{Sylvain Chaty} 
\author[esa]{Erik Kuulkers}
\author[iki]{Alexander Lutovinov}
\author[oas]{Angela Malizia}
\author[oas,unab]{Nicola Masetti}
\author[iki]{Ilya A. Mereminskiy}
\author[iki,mpa]{Rashid Sunyaev}
\author[utu,iki]{Sergey S. Tsygankov}
\author[inaf]{Pietro Ubertini}
\author[esa]{Christoph Winkler}


\begin{abstract}
The \integral\ hard X-ray surveys have proven to be of fundamental importance. \integral\ has mapped the Galactic plane with its large field of view and excellent sensitivity. Such hard X-ray snapshots of the whole Milky Way on a time scale of a year are beyond the capabilities of past and current narrow-FOV grazing incidence X-ray telescopes. By expanding the \integral\ X-ray survey into shorter timescales, a productive search for transient X-ray emitters was made possible. In more than fifteen years of operation, the \integral\ observatory has given us a sharper view of the hard X-ray sky, and provided the triggers for many follow-up campaigns from radio frequencies to gamma-rays. In addition to conducting a census of hard X-ray sources across the entire sky, \integral\ has carried out, through Earth occultation maneuvers, unique observations of the large-scale cosmic X-ray background, which will without question be included in the annals of X-ray astronomy as one of the mission's most salient contribution to our understanding of the hard X-ray sky.
\end{abstract}

\begin{keyword}
\end{keyword}

\end{frontmatter}


\tableofcontents

\section{Introduction}
\label{sec:intro}

A wide variety of astrophysical phenomena cannot be sufficiently well investigated via observations of individual sources, but requires instead a systematic approach based on large statistical samples. The last few decades of X-ray astronomy have provided us with great opportunities for studies of the populations of compact X-ray sources  (white dwarfs, neutron stars, black holes) in our Galaxy and beyond, with the use of new long-lasting facilities.

Two powerful and currently active hard X-ray missions, ESA's \integral\ observatory \citep{2003A&A...411L...1W} and NASA's Neil Gehrels Swift Observatory \citep{2004ApJ...611.1005G} are performing some of the deepest and widest serendipitous X-ray surveys ever undertaken at energies $E>20$~keV. In contrast to {\it Swift}, with a nearly uniform all-sky survey, which is especially useful for studies of active galactic nuclei \citep[AGN;][]{2010ApJS..186..378T,2010A&A...524A..64C,2012ApJ...749...21A,2013ApJS..207...19B,2018ApJS..235....4O}, the \integral\ observatory provides a sky survey with exposures that are deeper in the Galactic plane (GP) and Galactic Centre (GC) regions and with higher angular resolution, which is essential in these crowded regions. It allowed to study in depth different populations of galactic binary systems, such as low- and high-mass X-ray binaries, cataclysmic variables, symbiotic systems, etc. \citep[see, e.g.,][]{2008A&A...491..209R,2012ApJ...744..108B,2005A&A...444..821L,2013MNRAS.431..327L,2019NewAR..8601546K}. This makes the {\it Swift} and \integral\ surveys complementary to each other. In this review we concentrate on the valuable contribution of the \integral\ observatory to the surveying of the hard X-ray sky over the last 15 years.

The \integral\ observatory, selected as the M2 mission within ESA’s Horizon 2000 program, has been successfully operating in orbit since its launch in 2002. Due to the high sensitivity and relatively good angular resolution of its instruments, in particular the coded-mask telescope IBIS (Ubertini et al. 2003), surveying the sky in hard X-rays is one of the mission primary goals.


\begin{figure}
\begin{minipage}{0.49\textwidth}
\includegraphics[trim= 0mm 0cm 0mm 0cm,
  width=1\textwidth,clip=t,angle=0.,scale=0.98]{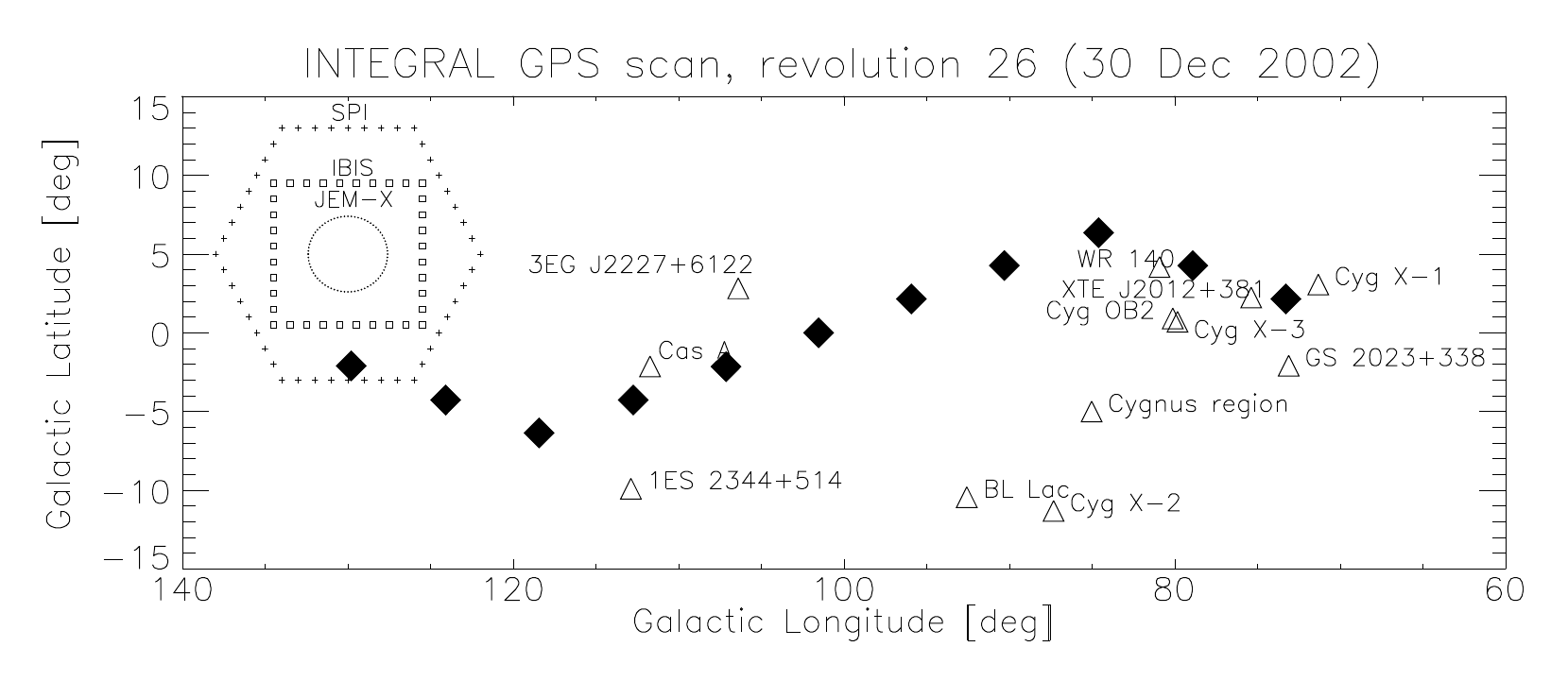}
\end{minipage}
\caption{Typical observational strategy of the \integral\ GPS scans (filled symbols). The fully coded fields of view (FCFOV, SPI: $16^{\circ}$, IBIS: $9^{\circ} \times 9^{\circ}$, JEM-X: $5^{\circ}$), and some known high energy targets are shown for illustration. Adopted from \cite{winkler2003}.
\label{fig:gps}
}
\end{figure}

\section{\integral\ hard X-ray surveys}

\subsection{Observations before \integral}

Since the beginning of X-ray astronomy many X-ray surveys have been successfully carried out with the aim of both discovering new types of X-ray emitters and to investigate the nature of the Cosmic X-Ray Background. A brief review of the hard X-ray surveys before the \integral\ era is presented hereafter. A more detailed overview of the Hard X-ray/Soft gamma-ray experiments and missions can be found in \cite{2017SSRv..212..429C}.


\cite{1979ApJS...39..573M} described observations of the cosmic X-ray sky performed by the MIT
1--40~keV X-ray detectors on the OSO~7 satellite between 1971 October and 1973 May. The authors made intensity determinations or upper limits for 3rd Uhuru \citep{1974ApJS...27...37G} and OSO~7 \citep{1976ApJ...206..265M,1977ApJ...218..801M} cataloged sources in different energy bands, including the 15--40~keV range.

{\it Uhuru}, also known as the Small Astronomical Satellite 1 (SAS-1) provided the first comprehensive and uniform all sky survey with a sensitivity of $10^{-3}$ the Crab intensity. \cite{1978ApJS...38..357F} presented the list of detected 339 X-ray sources with measured 2--6~keV intensities. The major classes of identified objects included binary stellar systems, supernova remnants, Seyfert galaxies and clusters of galaxies.

The first ever attempt to survey the sky at high energies ($26-1200$~keV) was performed in 1974-1979 with the Sky Survey Instrument on \textit{Ariel V} \citep{1982MNRAS.200..385C}, which provided the first galactic Log$N$-Log$S$ relation above 100~keV.

\cite{1987Natur.330..544S} reported observations of the Galactic Centre made with a coded mask X-ray telescope flown on the Spacelab 2 mission, providing the first images of the GC in high-energy X-rays up to 30~keV.


In the hard X-ray range ($2.8-30$~keV), an all-sky survey was conducted with the
BeppoSAX Wide Field Camera \citep[WFC;][]{1997A&AS..125..557J}. The WFC discovered 21 transients in the GC region and more than 50 transient and recurrent sources along the Galactic plane.

\cite{1984ApJS...54..581L} presented the first systematic study of X-ray sources at high X-ray energies (13--180~keV) over the whole sky. This all-sky survey was based on data obtained with the UCSD/MIT Hard X-Ray and Low-Energy Gamma-Ray Instrument A4 on board the HEAO~1 satellite from August 1977 until January 1979. The survey catalogue contains 72 sources at a flux sensitivity of $\sim10-15$~mCrab.

The Galactic Center region was observed with the TTM/COMIS coded mask imaging spectrometer on the Kvant module of the MIR orbital station in 1989 \citep{1991SvAL...17...54S,1991PAZh...17..126S}. Several observations of Galactic sources were performed, deriving the hard X--ray component of their emission \citep{1988PAZh...14..771S,1989SvAL...15..125S,1991SvAL...17..409S,1997ESASP.382..299B,1997AdSpR..19...29K}. \cite{1989ESASP.296..633S} presented the hard X-ray (2--30~keV) observations of the Large Magellanic Cloud (LMC) performed in 1988-1989 with the TTM/COMIS instrument, reporting the results  of monitoring and spectral observations of LMC~X--1, LMC~X--2, LMC~X--3, LMC~X--4 and PSR~0540-693. \cite{2000AstL...26..297E} assembled 
a catalog of 67 X-ray sources observed by the TTM/COMIS telescope in 1988-1998 at a confidence level higher than $4\sigma$. 

In 1990-1992 more than 400 sky fields were observed with the ART-P coded-mask telescope aboard the {\it GRANAT} observatory in the 2.5--60~keV energy band \citep{1990AdSpR..10b.233S}. ART-P provided a good $5'$ FWHM angular resolution within $3.4^{\circ}\times3.6^{\circ}$ field of view (FOV), which made it especially useful for studying the crowded field of the GC.  \cite{1994ApJ...425..110P} reported a detection of 12 point X-ray sources during a $\sim5^{\circ}\times5^{\circ}$ survey of the GC with the sensitivity of $\sim1$~mCrab in the $3-17$~keV energy range.

At higher energies, in the period of 1990--1998 the SIGMA telescope on board {\it GRANAT} observed more that one quarter of the sky with the sensitivity better than 100 mCrab \citep{2004AstL...30..527R}. The SIGMA telescope \citep{1991AdSpR..11h.289P}, designed in a coded-mask paradigm, allowed to reconstruct first images of the hard X-ray sky in the energy band $35-1300$~ keV with an angular resolution of $\sim15'$. During its operation time the SIGMA telescope detected 37 hard X-ray sources in the 40--100~keV energy band \citep{2004AstL...30..527R}. 


\subsection{\integral\ performance for surveys}

The IBIS telescope \citep{ibis} is the most suitable for the imaging surveys in the hard X-ray band among the major instruments on board \integral. This instrument provides the best combination of field of view, sensitivity and angular resolution needed to conduct a wide-angle survey of the sky in a reasonable amount of time. This was optimised with the scientific goal to regularly monitor a large fraction of the Galactic plane and to discover most of the expected transient sources, whose existence was anticipated by X-ray missions like {\it BeppoSAX} and {\it RXTE}, operating at lower energies and/or coarser spatial resolutions in the '90. The low-energy detector layer, ISGRI \citep[INTEGRAL Soft Gamma Ray Imager;][]{isgri} is made of a pixelated $128\times128$ CdTe solid-state detector that views the sky through a coded aperture mask. IBIS generates images of the sky with a $12'$ FWHM resolution over a $28^{\circ}\times28^{\circ}$ field of view in the working energy range $15-1000$~keV. 

The IBIS telescope is designed in a coded-aperture imaging paradigm. The sky is projected onto the detector plane through the transparent and opaque elements of the mask mounted above the detector plane. The sky reconstruction is based on the deconvolution of the detector image with the known mask pattern \citep[see][]{Fenimore:81,1987Ap&SS.136..337S}. The standard IBIS/ISGRI analysis is presented in the paper by \cite{2003A&A...411L.223G}. 

Thanks to the coded-aperture design, the IBIS telescope incorporates a very large fully-coded FOV of $9^{\circ}\times9^{\circ}$ (all source radiation is modulated by the mask) and partially-coded FOV of $28^{\circ}\times28^{\circ}$ (only a fraction of source flux is modulated by the mask). In addition to that, the ``dithering'' pattern around the nominal target position, a controlled and systematic spacecraft dithering manoeuvres introduced in order to minimize systematic effects due to spatial and temporal background variations in the spectrometer's (SPI) detectors, results in an even larger sky coverage. The combination of the standard $5\times5$ dithering grid and numerous \integral\ pointings, via the approved Guest Observer Program at the Galactic X-ray sources makes the effective latitude coverage of the Galactic plane $|b|<17.5^{\circ}$ \citep{krivonos2012}. As a result, \integral\ can conduct time-resolved mapping of the Galactic plane on the time-scale of a year. This leads to the unique possibility of taking snapshots of the whole Milky Way in hard X-rays, which is not possible with narrow-FOV grazing X-ray telescopes.


The energy response of the IBIS telescope at hard X-rays ($E>20$~keV) opens another possibility: to detect highly obscured objects. This makes the \integral/IBIS survey of the Galaxy unbiased against line-of-sight (or intrinsic to the source) attenuation of X-ray photons. Going to even higher energies, the ISGRI detector of \integral/IBIS can provide a census of Galactic hard X-ray emitters at energies above 100~keV \citep{2006ApJ...649L...9B,2015MNRAS.448.3766K}.

\subsection{Brief overview of INTEGRAL surveys}

\begin{figure*}[ht]
\centering
\begin{minipage}{0.75\textwidth}
\includegraphics[trim= 0mm 0cm 0mm 0cm,
  width=\textwidth,clip=t,angle=0.,scale=0.98]{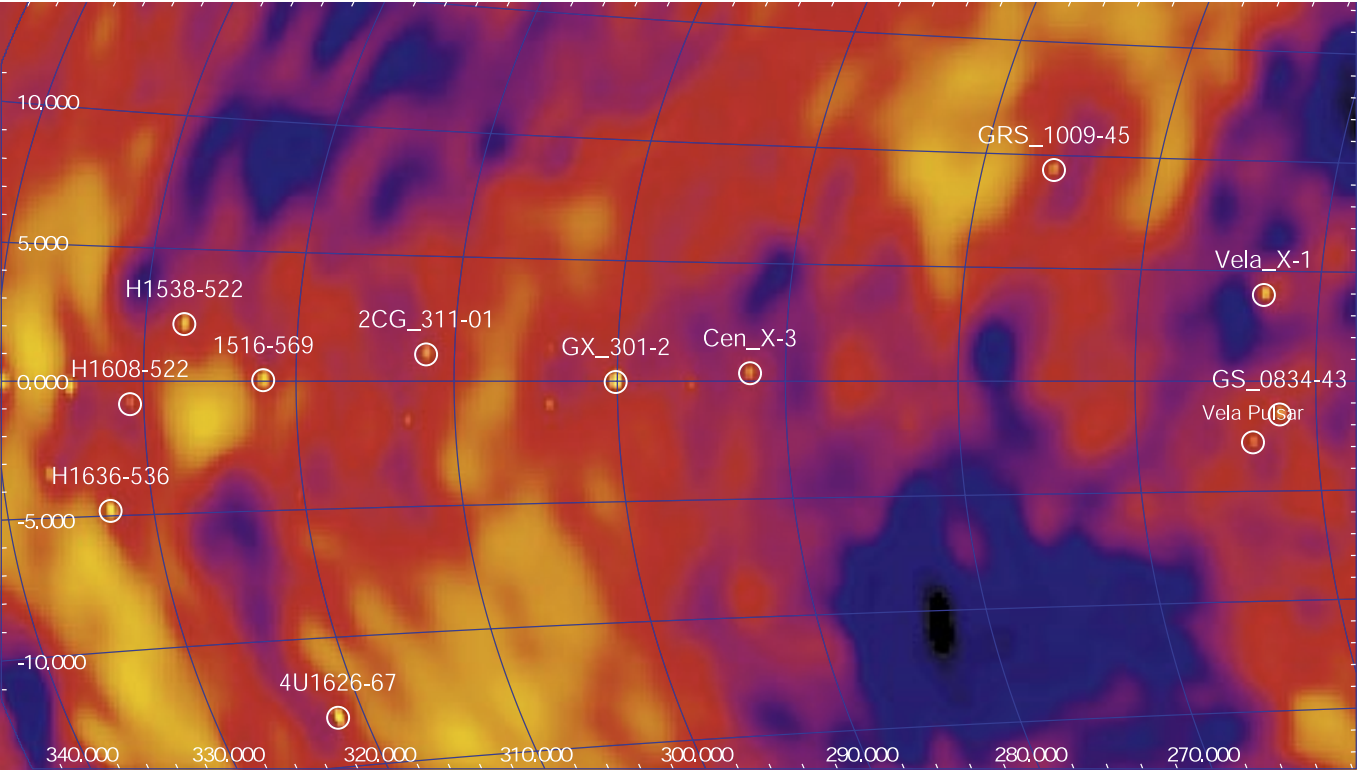}
\end{minipage}
\begin{minipage}{0.75\textwidth}
\includegraphics[trim= 0mm 0cm 0mm 0cm,
  width=\textwidth,clip=t,angle=0.,scale=0.98]{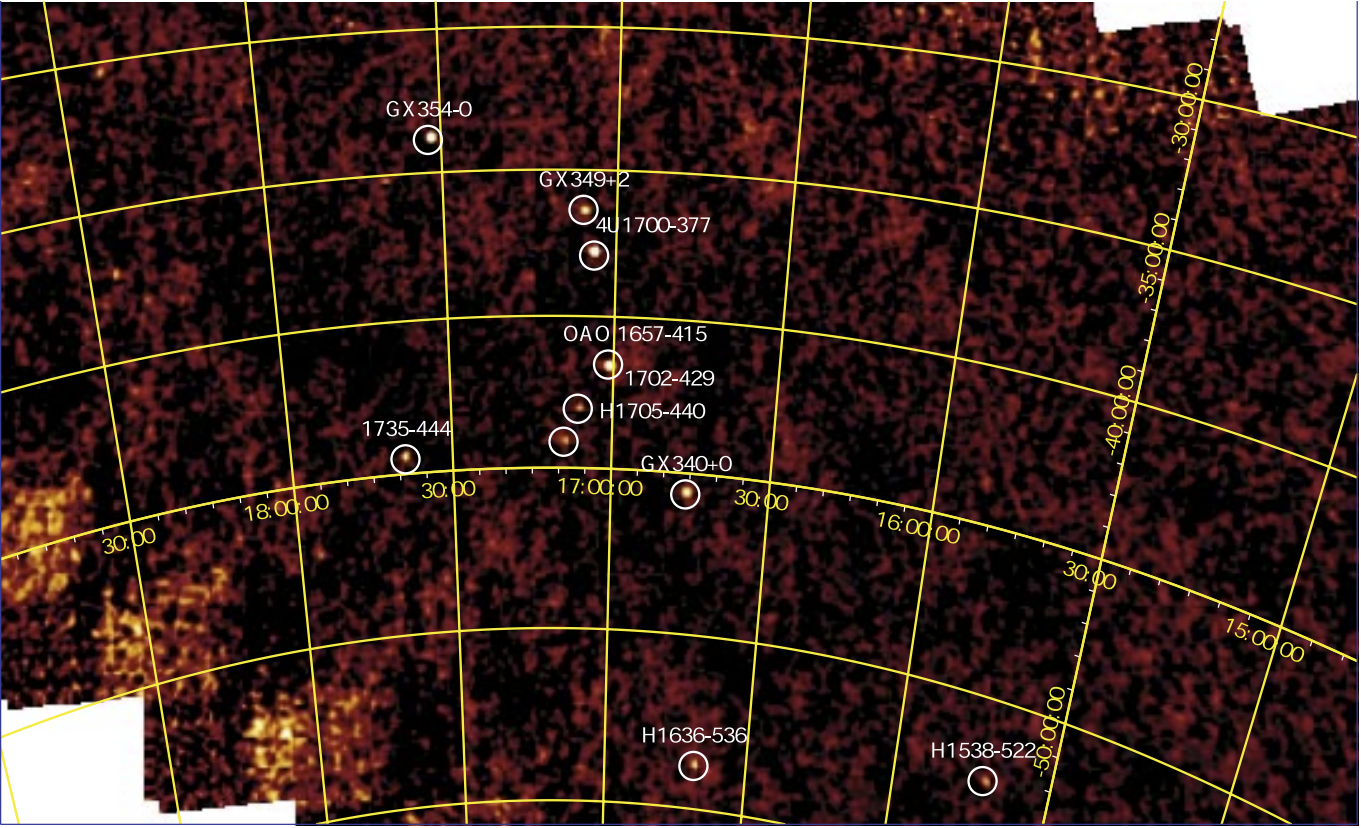}
\end{minipage}
\begin{minipage}{0.75\textwidth}
\includegraphics[trim= 0mm 0cm 0mm 0cm,
  width=\textwidth,clip=t,angle=0.,scale=0.98]{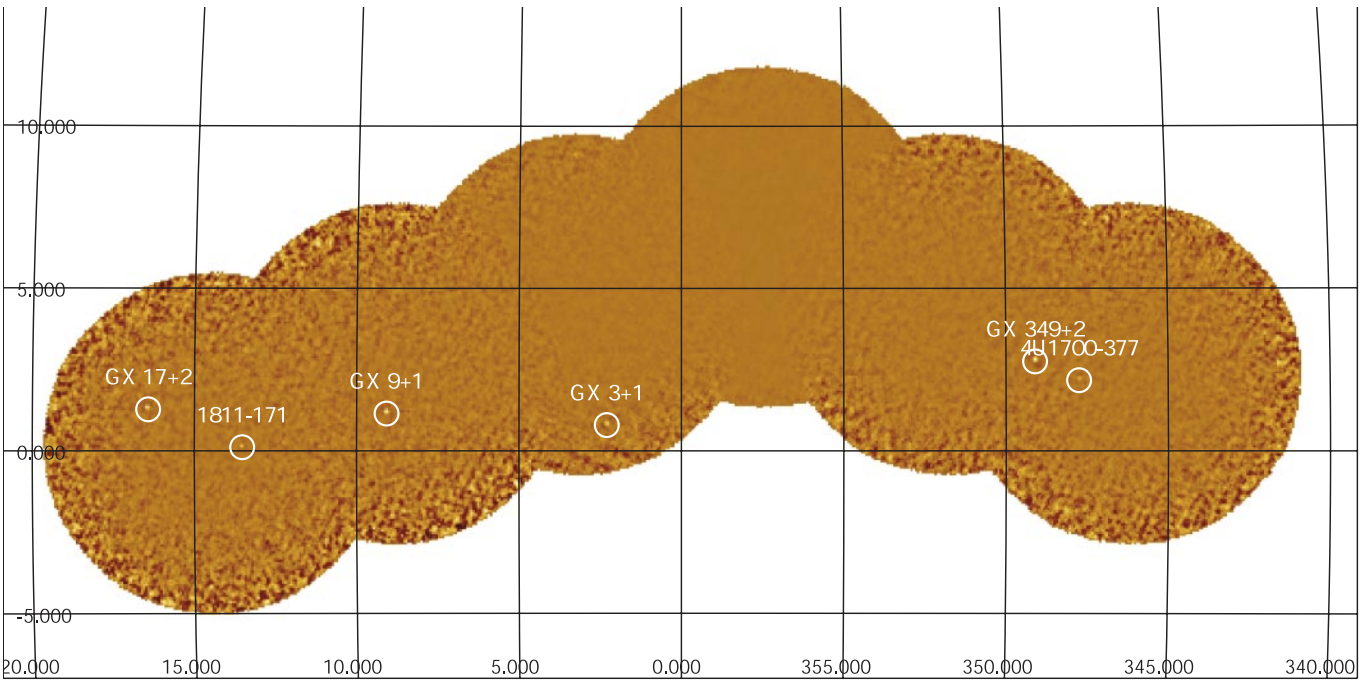}
\end{minipage}
\caption{Examples of Galactic sky maps obtained during the GPS observations. {\it Top panel}: SPI significance $20-40$~keV sky map in galactic coordinates (December 2002 -- March 2003). {\it Middle panel}: $15-30$~keV \ibis\ skymap in equatorial J2000 coordinates (12 March 2003). {\it Bottom panel}: $3-15$~keV \jemx\ sky map close to the Galactic Centre (24 March 2003). Adopted from \cite{winkler2003}.\label{fig:winkler2003:spi}
}
\end{figure*}

During the first years of operations, \integral\ conducted its so-called Core Programme (CP), i.e., a set of guaranteed observations dedicated to frequent monitoring of the Galactic plane in order to detect transient sources, and to perform time resolved mapping of the Galactic plane  \citep{2001ESASP.459..471W}. The CP consisted of a deep exposure of the central Galactic radian (GCDE: Galactic Centre Deep Exposure), regular scans of the Galactic plane (GPS: Galactic Plane Scan, Fig.~\ref{fig:gps}) and pointed observations. This program resulted in a first deep survey of the Galactic plane \citep{winkler2003}, which listed 100 known X-ray emitters and 10 new X-ray sources have been detected with IBIS/ISGRI for the first time. Fig.~\ref{fig:winkler2003:spi} shows the first  \integral\ wide-angle Galactic sky maps obtained during the GPS scans with SPI, IBIS/ISGRI, and JEM-X. During the first year of observations, \integral\ demonstrated its strength in discovering new transient sources. 10 new sources were found over a period of about 4~months, giving a discovery rate of more than 2 per month.

Deep surveys of the Galactic center region \citep{2004AstL...30..382R}, made within the Russian quota time, were a valuable addition to the \integral\ CP and demonstrated the excellent capabilities of the \integral\ telescopes to construct sensitive sky maps. In total, 60 sources with a flux higher than 1.5~mCrab were detected, including 2 sources detected at energies above 20~keV for the first time. 

Soon after, the \integral\ hard X-ray cartography of the Galactic plane has been extended by deep observations of the Sagittarius spiral arm \citep{2004AstL...30..534M}, a region very rich of young massive stars and remnants of their evolution. This part of the Milky Way contains the brightest well-known microquasar GRS~1915+105 \citep{2003AA...411L.415H}, the peculiar object SS433 \citep{cherss433}, soft gamma-ray repeaters, supernova remnants, about dozen of persistent and transient X-ray bursters (e.g. 4U~1915-05, Ser~X-1, Aql~X-1), and X-ray pulsars. As a result of the \integral\ Sagittarius arm survey, \cite{2004AstL...30..534M} reported the significant detection of 28~X-ray sources with a flux level above 1.4~mCrab, including three previously unknown X-ray emitters. 

Next step in the deep surveys of galactic spiral arms was made in 2005 with observations of the Galactic plane region in Crux \citep{2006AstL...32..145R}. In total it detected 47 hard X-ray sources, with 15 of them being new ones. Among the identified sources there were 12 active  galactic  nuclei, and 11 and 6  galactic  binary  systems  with  high-mass  and  low-mass  optical  companions,  respectively.   

In February 2005 the program of monitoring the source activity in the Galactic bulge region started\footnote{\url{http://isdc.unige.ch/Science/BULGE/}}. \cite{kuulkers2007} presented a detailed study of a homogeneous (hard) X-ray sample of 76~sources in the Galactic bulge region, based on the first 1.5~years of the program. The authors showed that almost all the sources in the Galactic bulge are variable. Additionally, 6 new hard X-ray sources were discovered.

Also, \cite{denHartog2006} presented results based on 1.6~Ms \integral\ observations of the Cassiopeia region. The analysis of the IBIS/ISGRI data resulted in detection of 11 sources at energies above 20~keV, including 3 new hard X-ray sources.

\onecolumn
\twocolumn

\noindent The discovery of hard X-ray emission from the anomalous X-ray pulsar 4U~0142+61 up to 150~keV demonstrated the unique capabilities of the IBIS/ISGRI imager.

Thanks to the continuing observations and, consequently, rapidly growing exposure, \cite{2004ApJ...607L..33B} released a first systematic survey of hard X-ray sources detected with the IBIS telescope, based on 5~Ms of total exposure time. This initial survey has revealed the presence of $\sim 120$ sources detected with the unprecedented sensitivity of $\sim 1$~mCrab in the energy range $20-100$~keV. The survey contains also 28~objects of unknown nature.

A sequence of IBIS/ISGRI survey catalogues has been released at regular basis as more data have become available. The second \integral/IBIS/ISGRI catalogue \citep{2006ApJ...636..765B} used a greatly increased data set of 10~Ms to unveil a soft gamma-ray sky comprising 209~sources, again with a substantial component (25\%) of new and unidentified sources.

\begin{figure*}[ht]
\begin{minipage}{0.64\textwidth}
\includegraphics[trim= 0mm 0cm 0mm 0cm,
  width=\textwidth,clip=t,angle=0.,scale=0.98]{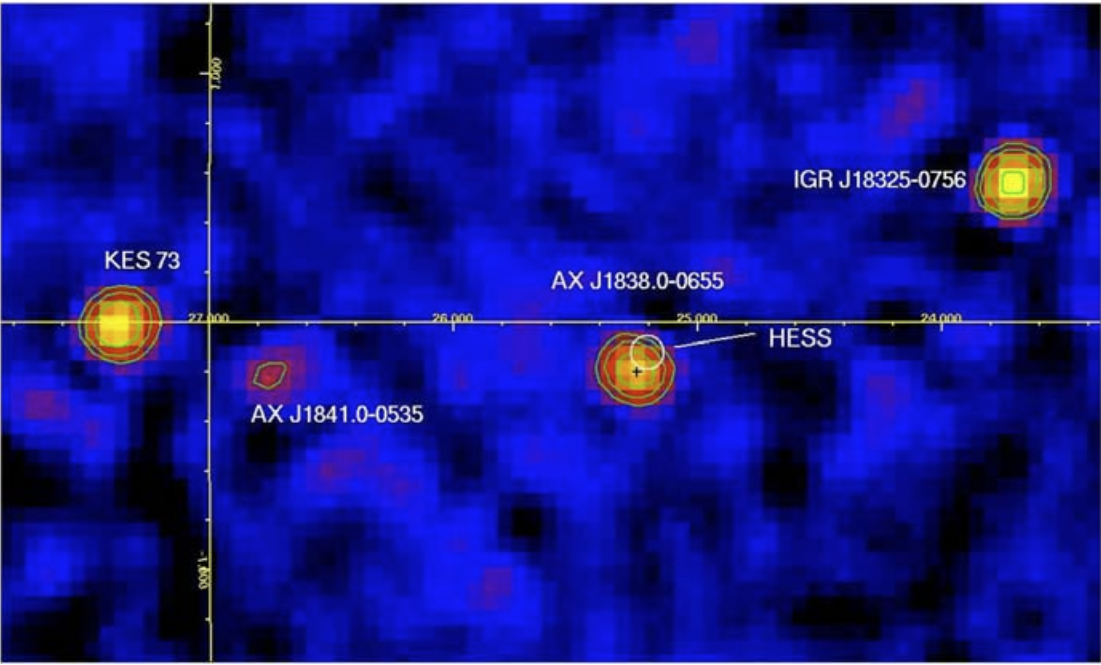}
\end{minipage}
\begin{minipage}{0.36\textwidth}
\includegraphics[trim= 0mm 0cm 0mm 0cm,
  width=\textwidth,clip=t,angle=0.,scale=0.98]{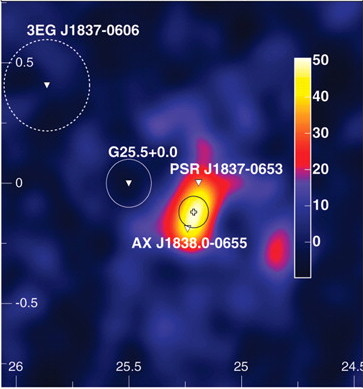}
\end{minipage}
\caption{\textit{Left:} The IBIS $20-300$~keV sky region map containing AX~J1838.0$-$0655 as well as the position and extension of HESS~J1837$-$069 (white circle) and the position determined by the {\it Einstein} telescope (black cross). Figure adapted from \cite{2005ExA....20...75U} and \cite{2005ApJ...630L.157M}. \textit{Right panel:} The emission region of HESS~J1837-069 overlapped to the ASCA source AX~J1838.0$-$0655 error box. Note that even if the two sources were positionally coincident, the TeV emission was suggestive of an extended object, confirmed at a later stage. Figure adapted from \cite{2005Sci...307.1938A}.
\label{fig:integral-hess}
}
\end{figure*}

In the meantime the first sensitive survey of the central part of the Galaxy was performed at very high energy gamma-rays  (E $>100$~GeV) with the {\it High Energy Stereoscopic System} \citep[HESS, ][]{2005Sci...307.1938A}. This survey revealed a new population of previously unknown sources emitting at very high energies. At least two had no known radio or X-ray counterpart. One year later the extended HESS survey of the Galaxy ($\pm30^{\circ}$ in longitude and $\pm3^{\circ}$ in latitude) confirmed the detection of 14 galactic sources \citep{2006ApJ...636..777A}.  Most of them had no known radio or X-ray counterpart and were hypothesised to be representative of a new class of dark nucleonic cosmic sources. In fact, high energy gamma-rays with energies $E>10^{11}$~eV were, and actually are, the best proof of non-thermal processes in the universe and provide a direct in-site view of matter-radiation interaction at energies by far greater than producible in human-made accelerators. \integral\ survey results were a powerful tool to immediately investigate the nature of this new type of galactic cosmic accelerator \citep{2005ExA....20...75U} thanks to the unprecedented \integral/IBIS angular resolution and Point Source Location Capability of about 1-2~arcminute between 15~keV and a few MeV. Furthermore, \integral\ had already performed deep observations of almost all HESS detected sources: the new INTEGRAL source IGR~J18135$-$1751 was identified as the soft gamma-ray counterpart of HESS~J1813$-$178 \citep[][]{2005ApJ...629L.109U} and AX~J1838.0$-$0655 as the X/gamma-ray counterpart of HESS~J1837$-$069 \citep{2005ApJ...630L.157M}, as shown in Fig.~\ref{fig:integral-hess}. It was immediately obvious that most of the common \integral/{\it HESS} sources identified belonged to SNR and Pulsar Wind Nebulae.

With the aim to provide a prompt release of the \integral\ hard X-ray survey information to the community, \cite{2007ApJS..170..175B} released a third \integral\ survey that covers $>70\%$ of the sky with an exposure of at least 10~ks. This ``all-sky'' survey, based on more than 40~Ms of \integral\ exposure, comprises more than 400 high-energy sources detected in the energy range $17-100$~keV, including both transients and faint persistent objects revealed on time-averaged maps.

Using alternative IBIS/ISGRI sky reconstruction software, \cite{krivonos2007a} released the first \integral/IBIS/ISGRI survey with all-sky coverage. The catalogue of detected sources includes 403~objects, 316 of which exceed a $5\sigma$ detection threshold on the time-averaged map of the sky, with the rest detected in various subsamples of exposures. Since this was the first all-sky survey in hard X-ray energy band with unprecedented sensitivity \citep[a factor of 10 deeper than the extragalactic part of the HEAO 1 A-4 source catalog, ][]{1984ApJS...54..581L}, the statistical properties of extragalactic X-ray source population (mainly AGNs) have been significantly improved. In particular, using 68 AGNs detected by \integral,  \cite{krivonos2007a} presented evidence of a strong inhomogeneity in the spatial distribution of nearby ($<70$~Mpc) AGNs, which reflects the large-scale structure in the local Universe. This finding has been later confirmed and significantly improved with $\sim6$ times larger AGN sample detected in {\it Swift}/BAT all-sky survey by \cite{2012ApJ...749...21A}.

\begin{figure*}[ht]
\begin{minipage}{\textwidth}
\centering
\includegraphics[trim= 0mm 0cm 0mm 0cm,
  width=\textwidth,clip=t,angle=0.,scale=0.98]{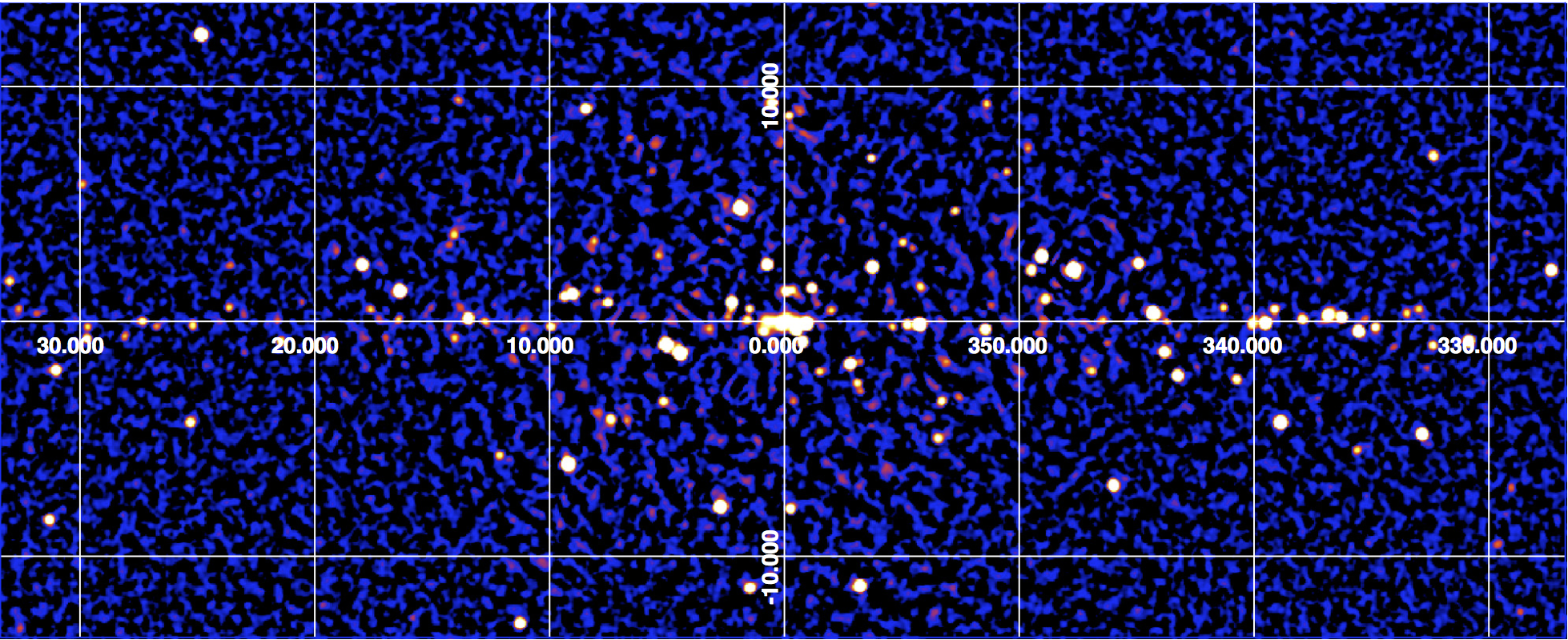}
\end{minipage}
\caption{Map of the central part of the Galaxy obtained with IBIS/ISGRI in the $17-60$~keV energy band using the improved sky reconstruction method \citep{2010A&A...519A.107K}. The total exposure is about 20~Ms in the region of the Galactic center. To highlight background fluctuations, the image is shown in significance with a squared root color map ranging from 0 to 25. As a consequence of the chosen color scheme, the apparent diameters of the source images partially scale as the X-ray brightness of the sources. Figure adapted from \cite{2010A&A...519A.107K}.
\label{fig:krivonos2010-gc}
}
\end{figure*}

In the following years, \integral\ continued to accumulate exposure time within the Galactic plane. However, the growing exposure time was not reflected by a corresponding increase in survey sensitivity, since the observations are strongly affected by the systematics related to the crowded field of the Galactic center. \cite{2010A&A...519A.107K} developed an improved method of sky image reconstruction for the IBIS telescope, which allowed them to significantly suppress the systematic noise in the deep images of the Galactic center (see Fig.~\ref{fig:krivonos2010-gc}), and practically remove non-statistical noise from the high-latitude sky images. This improved method of sky reconstruction was used by \cite{krivonos2010b} to conduct the most sensitive survey of the Milky Way above 20~keV at that time. The minimal detectable flux with a $5\sigma$ detection level reached the value of $3.7 \times 10^{-12}$\ergscm, which corresponds to $\sim 0.26$ mCrab in the $17-60$~keV energy band. The catalogue of detected sources includes 521~objects, 449 of which exceed a $5\sigma$ detection threshold on the time-averaged map of the sky, and 53 were detected in different periods of observations. Among the identified sources, 262 are Galactic and 221 are of extragalactic origin.

\cite{2010ApJS..186....1B} presented the fourth soft gamma-ray source catalogue obtained with IBIS/ISGRI based on 70~Ms of high-quality observations performed during the first five and a half years of the Core Program and public observations. The catalogue includes a substantially increased coverage of extragalactic fields, and comprises more than 700 high-energy sources detected in $17-100$~keV energy range. The authors performed careful analysis of IBIS data using the latest official OSA software\footnote{The \integral\ Off-line Scientific Analysis (OSA) package is provided by \integral\ Science Data Center (ISDC, \citet{2003A&A...411L..53C}) to the community to reduce and analyze data collected by the \integral\ satellite.} and source detection techniques. Particular care has been taken to optimize the detection of the transient sources that are common to find both transients and faint persistent objects that can only be revealed with longer exposure times.

Six years later, \cite{2016ApJS..223...15B} reported an all-sky catalogue of soft gamma-ray sources based on IBIS observations during the first 1000~orbits of \integral. This legacy-level survey contains all good-quality data acquired from the launch in 2002, up to the end of 2010 and contains 110~Ms of scientific public observations, with a concentrated coverage on the Galactic Plane and extragalactic deep exposures. The catalogue includes 939 sources above a $4.5\sigma$ significance threshold detected in the $17-100$~keV energy band. The list of previously undiscovered soft gamma-ray emitters contains 120 sources. Substantial efforts have been taken to detect transient sources on different time scales as described in Section~\ref{sec:bursticity}.

\begin{figure}[ht]
\begin{minipage}{\columnwidth}
\centering
\includegraphics[trim= 0mm 0cm 0mm 0cm,
  width=\textwidth,clip=t,angle=0.,scale=0.98]{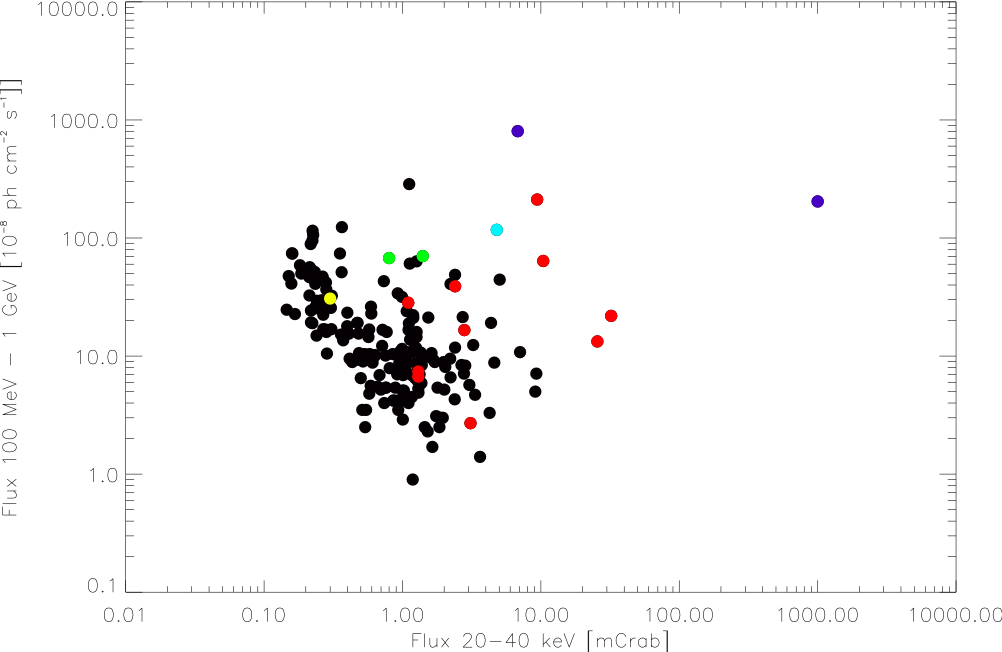}
\end{minipage}
\caption{Gamma-ray flux (100 MeV -- 1 GeV) of each {\it Fermi}/LAT source as a function of the corresponding 20--40 keV IBIS/ISGRI flux. The coloured points refer to the IBIS detections, specifically red points are blazars, dark blue are pulsars, green are HMXBs, yellow is Eta Carinae and finally light blue is IGR~J17459--2902. The black points refer to IBIS non-detections (2$\sigma$ upper limit). Figure adapted from \cite{2009ApJ...706L...7U}.
\label{fig:ubertini2009}
}
\end{figure}

In June 2008 the {\it Fermi Gamma-ray Space Telescope} was successfully launched and soon after the first high energy catalogue was published. \cite{2009ApJ...706L...7U} reported the result of the cross correlation between the 4th \integral/IBIS soft gamma-ray catalogue \cite{2010ApJS..186....1B}, in the range 20--100~keV, and the {\it Fermi}/LAT bright source list of objects emitting in the 100~MeV -- 100~GeV range. Surprisingly, the main result was that only a minuscule part of the 720 sources detected by INTEGRAL were present among the 205 {\it Fermi}/LAT sources (Fig.~\ref{fig:ubertini2009}). This result was not expected due to the mCrab \integral\ sensitivity and the {\it Fermi} breakthrough at MeV--GeV energies. Most of the {\it Fermi}/LAT gamma-ray sources present in the 4th INTEGRAL/IBIS catalogue were optically identified as AGNs (10) complemented by 2 isolated pulsars (Crab and Vela) and 2 High Mass X-Ray Binaries (HMXB, LS~I~+61$^{\circ}$303 and LS5039). Two more possible associations were found: one is 0FGL~J1045.6--5937, possibly the counterpart at high energy of the massive colliding wind binary system Eta Carinae, discovered to be a soft gamma-ray emitter. For the remaining 189 {\it Fermi}/LAT sources no \integral\ counterparts were found. 

This initial (unbiased) cross-correlation between low and high energy gamma-ray sources showed that MeV/GeV {\it Fermi} sources were usually not associated with IBIS/ISGRI sources in the range 20 -- 100 keV. The handful of objects common to both surveys comprised only Flat Spectrum Radio Quasars (FSRQ) and BL Lac objects, but no X-Ray Binaries, with the exception of the two microquasars mentioned. Also absent were the AXP, which were known to be strong emitters in the  keV-MeV range with a total energy rising in $\nu$F$_\nu$ and no cut-off detected in their spectra up to a few hundreds of keV \citep{2009AA...501.1031K}, implying some kind of switch-off mechanism in the MeV regime. Similarly, SGR and Magnetars, detected even in quiescence mode by IBIS/ISGRI \citep{2009MNRAS.396.2419R} and among the brightest sources of the hard X-ray sky when flaring \citep{1999ApJ...510L.115K,2006A&A...445..313G,2008ApJ...685.1114I} were not detected in the high energy gamma-ray range.

\begin{figure*}
\centering
\begin{minipage}{0.75\textwidth}
\includegraphics[trim= 0mm 0cm 0mm 0cm,
  width=1\textwidth,clip=t,angle=0.,scale=0.98]{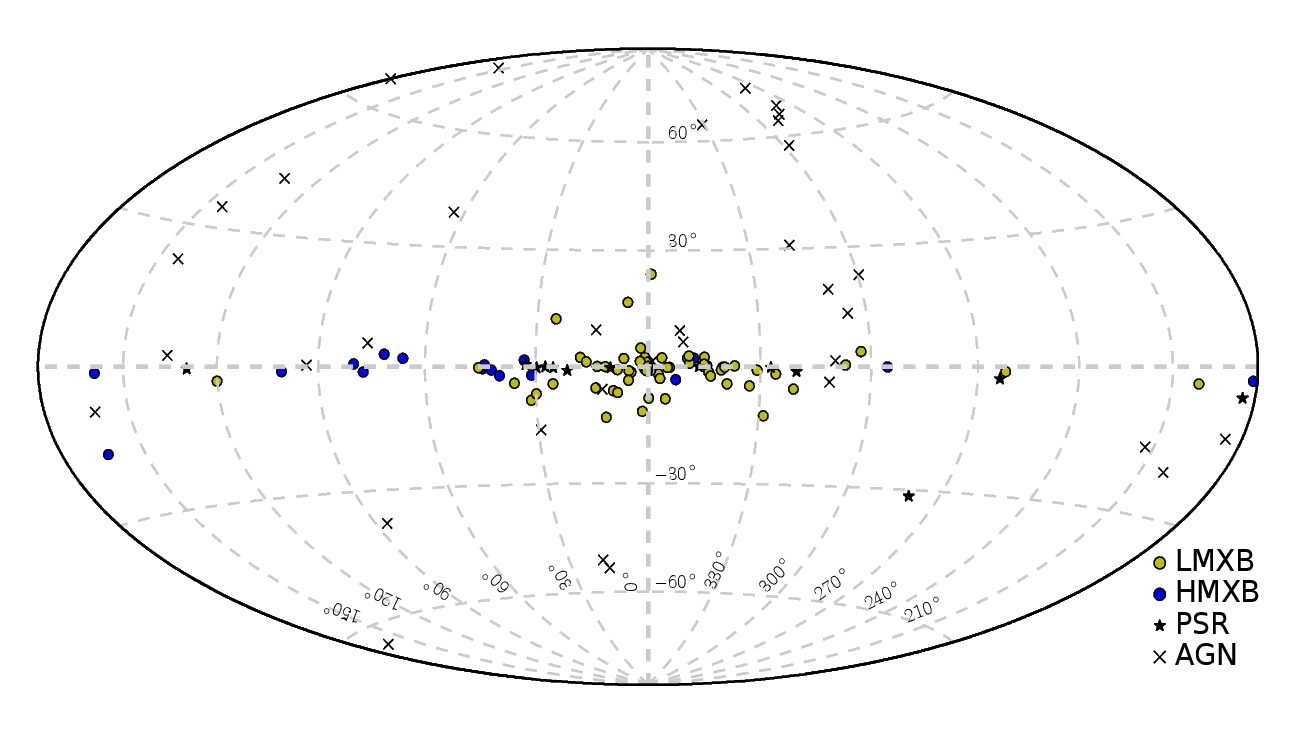}
\end{minipage}
\caption{All-sky map showing the four basic X-ray source types detected in the $100-150$~keV survey by \cite{2015MNRAS.448.3766K}: 65~LMXBs, 19~HMXBs, 13~PSRs (mostly in the Galactic plane) and 35~AGNs. Adapted from \cite{2015MNRAS.448.3766K}.
\label{fig:100keV}
}
\end{figure*}

As mentioned earlier, most of the observing time of \integral\ is dedicated to Galactic source population studies, making possible the deepest Galactic survey in hard X-rays ever compiled. Using sky reconstruction algorithms especially developed for the high quality imaging of IBIS/ISGRI data \citep{2010A&A...519A.107K}, \cite{krivonos2012} published an \integral\ Galactic plane survey based on nine years of observations, from December 2002 to January 2011. As seen from the range of the used spacecraft revolutions 26-1013, the time span of this survey is similar to that covered by 1000-orbits survey by \cite{2016ApJS..223...15B}. \cite{krivonos2012} presented sky images, sensitivity maps, and catalogues of detected sources in different energy bands energy bands ($17-60$, $17-35$, and $35-80$~keV) within the Galactic plane ($|b| < 17.5^{\circ}$). Using the extended data set, \cite{2017MNRAS.470..512K} reported on a catalogue of new hard X-ray source candidates based on the sky maps comprising 14~years of data acquired with the IBIS telescope within the Galactic Plane ($|b| < 17.5^{\circ}$). The catalogue includes in total 72 hard X-ray sources detected at $S/N > 4.7\sigma$ and not known to previous INTEGRAL surveys. 

Most of the \integral\ surveys have been conducted in the $17-100$~keV regime where the IBIS/ISGRI sensitivity is optimal in search for point sources. However, energy response of ISGRI detector allows to effectively detect photons at even higher energies, as seen from many studies of bright X-ray sources of different nature \citep[see e.g.,][]{2012MNRAS.423.1978L, 2014MNRAS.445.1205R, 2015ApJ...813L..21N, 2015ApJ...812...62C, 2016A&A...591A..66K, 2016MNRAS.458.2454L,2017A&A...603A..16D}. The first systematic study of X-ray emitters detected with IBIS/ISGRI in soft gamma-ray band $100-150$~keV has been conducted by \cite{2006ApJ...649L...9B} based on the Core Program and public open-time observations up to 2005 April. The catalogue includes 49 sources detected in the $100-150$~keV band, of which 14 are also seen in the $150-300$~keV range. The source types in $100-150$~keV band are dominated by Galactic low and high mass X-ray binary systems, and also include active galaxies (10). Among the binary systems that are detected above 150~keV, more than 50$\%$ are associated with black hole candidates. \cite{2006ApJ...649L...9B} constructed the first $100-150$~keV Galactic and extragalactic Log$N$-Log$S$ relation, predicting at $E>100$~keV around 200 Galactic sources and almost 350 active galaxies at a flux above 1~mCrab.

Ten years later, using significantly increased exposure time, \cite{2015MNRAS.448.3766K} published an \integral\ all-sky survey at energies above 100 keV. The catalogue of detected sources includes 132 objects, which significantly increases the high-energy source sample compared to the work of \cite{2006ApJ...649L...9B}. The whole sky map of all the detected sources in the survey, discriminated in four basic source classes, is shown in Fig.~\ref{fig:100keV}. The survey is dominated by 97 hard X-ray sources of Galactic origin (mainly Low-Mass X-ray Binaries -- LMXBs and HMXBs, 83 in total) in comparison with the extragalactic source population, represented by 35 AGNs. Compared to \cite{2006ApJ...649L...9B}, the Log$N$-Log$S$ was extended down to fainter fluxes by a factor of 1.4 and has a steeper slope. 

\integral\ regular observations of the Galactic plane make it possible to address non-standard questions: for instance, \cite{2016MNRAS.458.3411T} performed the deepest systematic search for the nuclear de-excitation lines of titanium-44 ($^{44}$Ti) at 67.9 and 78.4 keV, as a tracer of core-collapse supernova explosions in the Galaxy. The peak sensitivity of this $^{44}$Ti survey reached an unprecedented level of $4.8\times10^{-6}$~ph~cm$^{-2}$~s$^{-1}$ that improved the sensitivity of the survey done by Compton Gamma-Ray Observatory/COMPTEL \citep{1999ApL&C..38..383I} by a factor of $\sim5$. As a result, constraining upper limits for all sources from the catalogue of Galactic supernova remnants \citep[SNR; ][]{2014BASI...42...47G} were derived. These upper limits can be used to estimate the exposure needed to detect $^{44}$Ti emission from any known SNR using existing and prospective X- and gamma-ray telescopes.

Thanks to large observational campaigns of the extragalactic sky, \integral\ accumulated a number of deep fields at high Galactic latitudes. The first \integral\ extragalactic survey was conducted in the direction toward the Coma cluster of galaxies by \cite{2005ApJ...625...89K}, who  detected 12 serendipitous sources with statistical significance $>4\sigma$ and extended the extragalactic source counts in the $20-50$~keV energy band down to a limiting flux of $\sim$1~mCrab. This is more than a factor of 10 improvement in sensitivity compared to the previous hard X-ray results in this energy band obtained with the HEAO 1 A-4 instrument. As a significant step forward, \cite{2006ApJ...652..126B} compiled a complete extragalactic sample based on a $\sim25,000$~deg$^2$ sky coverage down to a limiting flux of $3\times 10^{-11}$\ergscm~in $20-40$~keV. The sample of 38 AGNs was used to construct Log$N$-Log$S$ and to produce the first luminosity function of AGNs in the $20-40$~keV energy range. The census of nearby AGNs and their statistical properties was later extended by \cite{2007A&A...462...57S} using a representative sample of 127 AGN from \cite{krivonos2007a}. Later on, \cite{paltani2008} presented an analysis of a deep hard X-ray survey of the 3C 273/Coma region with sky coverage of about 2500 deg$^2$, resulting in a list of 34 candidate sources detected in the mosaic with a significance $\sigma > 5$. Another extragalactic field, that of the LMC, was scrutinized during the large observational campaign aimed at detecting the emission lines from the decay of $^{44}$Ti in the remnant of SN1987A \citep{2012Natur.490..373G}. The catalogue of sources in the LMC region was published by \cite{2013MNRAS.428...50G} and consisted of 21 sources, 4 of which were detected in hard X-rays for the first time. Later, a number of deep extragalactic fields, including M81, LMC and 3C 273/Coma, were studied in \cite{2016MNRAS.459..140M}, who detected 147 sources at $S/N > 4\sigma$, including 37 sources observed in hard X-rays for the first time.

The SPI spectrometer with its comparably large field of view provides the opportunity to expand the energy range of the \integral\ surveys up to a few MeV. Based on only the first year's data, \cite{2005ApJ...635.1103B} detected 63 sources at energies below 100~keV and four above 300~keV. The main contribution made by SPI was done in studies of the diffuse emission of the Galaxy, which are beyond the scope of the current review \citep[see review by][]{2014AstRv...9c...1D}. However, the positron annihilation line at 511~keV may have not only diffuse origin but can also originate from the very vicinity of compact objects. A systematic search for outbursts in the narrow positron annihilation line on various time scales based on the \integral/SPI data was performed by \cite{2010AstL...36..237T}. As a result, upper limits on the rate of outbursts with a given duration and flux in different parts of the sky were provided.

The two \jemx\ telescopes on-board \integral\ have a smaller (partially coded) field of view ($10^{\circ}$ in diameter) compared to the \ibis\ partially coded FOV of $28^{\circ} \times 28^{\circ}$, which restricted their ability to conduct wide-angle surveys; however, \cite{2015AstL...41..765G} released an X-ray survey of the GC region based on $\sim10$ years of observations (2003-2013). 

Table~\ref{tab:surveys} summarizes the \integral\ surveys conducted with IBIS/ISGRI, JEM-X, and SPI, listing important survey characteristics such as limiting flux, sky coverage, total number of detected sources and completeness. Since this table is sorted by year of publication, one can see how sensitivity improves with time, resulting in the growing number of detected sources. It is not trivial to count all hard X-ray sources discovered with \integral\ in different surveys by different research groups; however, as seen in Table~\ref{tab:surveys}, the total number of new IGR sources can easily reach several hundreds of objects, which demonstrates great impact of \integral\ in surveying the hard X-ray sky.  

\begin{table*}[t]
  \centering
  \begin{tabular}{lccrrrrrr}
    Paper by  & 
    \integral & 
    $\Delta E$ & 
    Sensitivity & 
    Sky &
    Total number &
    IGR  & 
    Completeness$^{d)}$ \\
    & 
    telescope & 
    [keV] & 
    [mCrab ($\sigma$)] & 
    coverage &
    of sources & 
    sources$^{c)}$ & &\\
  \hline
    \cite{winkler2003}  &  \ibis & $15-40$ &36$^{a)}$ ($5\sigma$)& & 110 & 10 & \\
                        & \spi & $20-40$   &62$^{a)}$ ($5\sigma$)& & 33 & 3  & \\
                        & \jemx & $5-20$   &20$^{a)}$ ($5\sigma$)& & 50 &  & \\
\cite{2004AstL...30..382R}  &  \ibis & $18-60$ & $1-2$ & $\sim 900$ deg$^{2}$ & 60 &  5 & 10/60 \\
\cite{2004AstL...30..534M} & \ibis & $18-60$ & $1.4$ & $35^{\circ}\times 40^{\circ}$ & 28 &  7 & 7/28 \\
\cite{2004ApJ...607L..33B} & \ibis & $20-100$ & $\sim1$ &  & 120 &  5 & 28/120 \\
\cite{2005ApJ...625...89K} & \ibis & $20-50$ & $\sim1$ ($4\sigma$) & $40^{\circ}\times 40^{\circ}$  & 13 & 5 & 5/13\\
\cite{2006AstL...32..145R} & \ibis & $17-60$ & $0.8-1$ ($5\sigma$) & $50^{\circ}\times 50^{\circ}$  & 46 & 20 & 13/46\\
\cite{2005ApJ...635.1103B} & \spi & $20-150$ & & $100^{\circ}\times 50^{\circ}$ & 63 & 8&\\
\cite{2006ApJ...636..765B}& \ibis & $20-100$ & $\sim1$ & $\sim50\%$  & 209 &  56 & $\sim75\%$ \\
\cite{2006ApJ...649L...9B} & \ibis & $100-150$ & $\sim2$ ($4\sigma$) & $\sim50\%$ & 49 & & $100\%$ \\
\cite{2007ApJS..170..175B} & \ibis & $17-100$ & $\sim1$ & $\sim70\%$  & 421 & 167 & $\sim75\%$ \\
\cite{krivonos2007a} & \ibis & $17-60$ & $\sim1$ & $100\%$  & 403 & 137 & 48/403 \\
\cite{kuulkers2007}   &  \ibis & $20-60$, &1$^{b)}$ ($3\sigma$)& & 76 & 18 & \\
                      &  & $60-150$       &3$^{b)}$ ($3\sigma$)& & 76 & 18 & \\
                      & \jemx & $3-10$, & & & 18 &  & \\
                      &  & $10-25$      & & & 18 &  & \\
\cite{paltani2008} & \ibis & $20-60$ & 0.5 ($5\sigma$) & 2500 deg$^2$ &  34 & 34 &  $\sim100\%$\\
\cite{krivonos2010b} & \ibis & $17-60$ & $0.26$ ($5\sigma$) & $100\%$  & 521 &  212 & 38/521 \\
\cite{2010ApJS..186....1B} & \ibis & $17-100$ & $<1$  & $100\%$  & 723 & 378  &  $\sim70\%$\\
\cite{krivonos2012} & \ibis & $17-80$ & $\sim0.2$ ($4.7\sigma$) & $|b| < 17.5^{\circ}$  & 402 & 180 & $\sim92\%$ \\
\cite{2013MNRAS.428...50G} & \ibis & $20-60$ & $\sim$0.5 ($4.5\sigma$) & 640 deg$^{2}$ & 21 & 4 & 90$\%$\\
                           & \jemx & $3-20$ &  $\sim$0.5 ($5\sigma$) & $\sim$100 deg$^{2}$ & 10 & 0 & 100$\%$\\
\cite{2015MNRAS.448.3766K} & \ibis & $100-150$ & $\sim2$ ($4\sigma$) & $100\%$ & 132 & & $100\%$ \\
Grebenev et al. (2015) & \jemx & $5-25$ & & $|l,b| < 20^{\circ}$ & 105 & 24& \\
\cite{2016ApJS..223...15B} & \ibis & $17-100$ & $<1$  & $100\%$  & 939 & $\sim560$  &  \\
\cite{2016MNRAS.458.3411T} & \ibis & $64.6-82.2$ & $\sim0.7$ ($4.7\sigma$) &  $|b| < 17.5^{\circ}$ & 1 & &\\
\cite{2016MNRAS.459..140M} & \ibis & $17-60$ & $\sim0.18$ ($4\sigma$) &  4900 deg$^2$ & 147 & 37 & 25/147\\
\cite{2017MNRAS.470..512K} & \ibis & $17-60$ & $\sim0.15$ ($4.7\sigma$) &  $|b| < 17.5^{\circ}$ & 72 & 72 & 46/72\\
  \end{tabular}
  \caption{The list of the \integral\ surveys.}
  \label{tab:surveys}
  \begin{flushleft}
  $^{a)}$ Average sensitivity per one GPS scan.
  $^{b)}$ Average sensitivity per season.
  $^{c)}$ The total number of IGR sources discovered with \integral\ in a given survey or previous works.
  $^{d)}$ The completeness column describes the fraction of sources with known nature, if specified with percentile. The numbers shown as a fraction represent the number of unclassified sources with respect to the total number of sources detected.
  \end{flushleft}
\end{table*}

\begin{figure}
\centering
\begin{minipage}{0.98\columnwidth}
\includegraphics[trim= 0mm 0cm 0mm 0cm,
  width=1\textwidth,clip=t,angle=0.,scale=0.98]{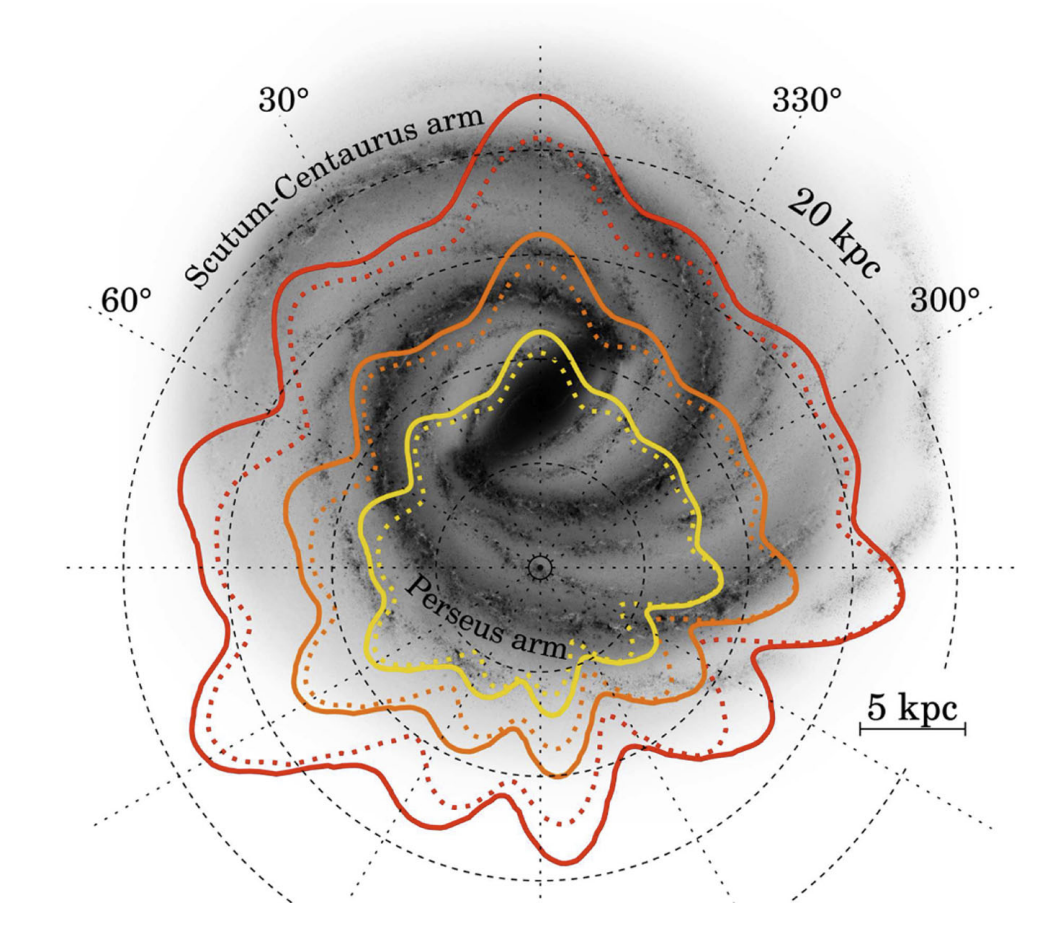}
\end{minipage}
\caption{Face-on view of the Galaxy shown along with the distance
  range at which an X-ray source of a given luminosity $L_{\rm HX}$
  (or more) can be detected according to the $17-60$~keV sensitivity
  of the 14-year \integral\ survey \citep[solid lines,][]{2017MNRAS.470..512K}, compared to the
  9-year Galactic plane survey \citep[dotted lines,][]{krivonos2012}. Red,
  orange and yellow contours correspond to $L_{\rm
    HX}=2\times10^{35}$, $10^{35}$ and $5\times10^{34}$~\lum,
  respectively.  The background image is a sketch of the Galaxy
  adapted from \citet{2009PASP..121..213C}.
\label{fig:galaxy}
}
\end{figure}

The on-going survey of the Galactic plane with \integral\ provides a continuous improvement in sensitivity, which makes it possible to probe deeper into the Galaxy. Fig.~\ref{fig:galaxy} shows a face-on schematic view of the Milky Way and the distances at which \integral\ can observe a hard X-ray source of a given luminosity $L_{\rm HX}$. One can notice that \integral\ can detect all sources with the luminosity $L_{\rm HX}>2\times10^{35}$~\lum at the far end of the Galaxy in the direction towards the Galactic Centre; the distance range for the luminosity $L_{\rm HX}>2\times10^{35}$~\lum covers most of the Galactic stellar mass; and the Galactic central bar is fully reachable at luminosities $L_{\rm HX}>5\times10^{34}$~\lum\ \citep{2017MNRAS.470..512K}.



\subsection{Time domain}
\label{sec:bursticity}

The deep sensitivity of modern hard X-ray surveys is largely achieved by stacking large numbers of relatively short exposures taken for the same fields over many years. In the case of IBIS, observations are divided into short pointings, or {\em science windows} of typically 2000~s, separated by short slews during which the instrument pointing direction changes by a few degrees. Each science window can be considered an independent measurement of the flux from all points in the field of view for that pointing.

The final outcome of this stacking  approach is essentially used to derive the weighted mean of many 2000~s of measurements of source flux taken over a time period in excess of a decade. The weighted mean is used because the measurement quality is non-uniform, being affected by several factors such as exposure time, changing position of the source in the field of view, and the presence of other bright sources in the field. For a {\em persistent} emitter, the weighted mean of the flux, and the error on that weighted mean, is an excellent estimator of the mean flux and how significantly the mean flux is non-zero; this is the detection significance usually quoted in survey catalogues. In other words, in the assumption that the source is persistent, the significance tells us how confident we can be that we detect a non-zero flux from a given sky position. But this assumption fails for variable or transient sources.

In order to optimise transient source detection, sources must be searched for on different timescales. This may be done with the construction of multiple maps covering different time periods, or directly on light curves of known (or suspected) sources. Inevitably, biases are introduced when we search for emission on a specific set of timescales, as we must make the problem tractable. 

The first IBIS survey \citep{2004ApJ...607L..33B} employed a straightforward stacking analysis, but from the second catalogue \citep{2006ApJ...636..765B} onward, which analysed $\sim$18 months of data, it was realised that source searches on additional time intervals would be needed to optimise the detection of sources that only emitted on shorter timescales. Consequently, maps were constructed and searched not only for the full archive, but also for each revolution (satellite orbit, $\sim$3 days) and for the periods covering the Galactic Centre Deep Exposure (GCDE) core programme ($\sim$1 month each). In the third catalogue \citep{2007ApJS..170..175B} the GCDE mosaics were replaced by a broader category of {\em revolution sequences} covering any observing period where the telescope performed a deep exposure on a single field. During the third catalogue construction, it was noticed that sources detected in previous catalogues were becoming difficult to detect, and strategies were developed to deal with the increasing baseline of the dataset. The problem is illustrated by the detection of the gamma-ray burst GRB041219A, one of the brightest sources ever detected by INTEGRAL. A strong detection of GRB041219A in the specific science window reduces rapidly as the observation window lengthens, and if more than $\sim$5 revolutions of data are stacked, the source is no longer detectable. The 4th catalogue \citep{2010ApJS..186....1B} introduced {\em bursticity} analysis, a sliding-window analysis that sought to detect sources on whatever timescale optimised their detection significance. Most recently, the bursticity method was refined for the catalogue of 1000 orbits \citep{2016ApJS..223...15B} which had a dataset spanning 8 years of satellite operations, and yet searches were performed for transient emission on timescales down to 0.5 days. 

\begin{figure}[htbp]
\resizebox{\hsize}{!}{\includegraphics[clip=true]{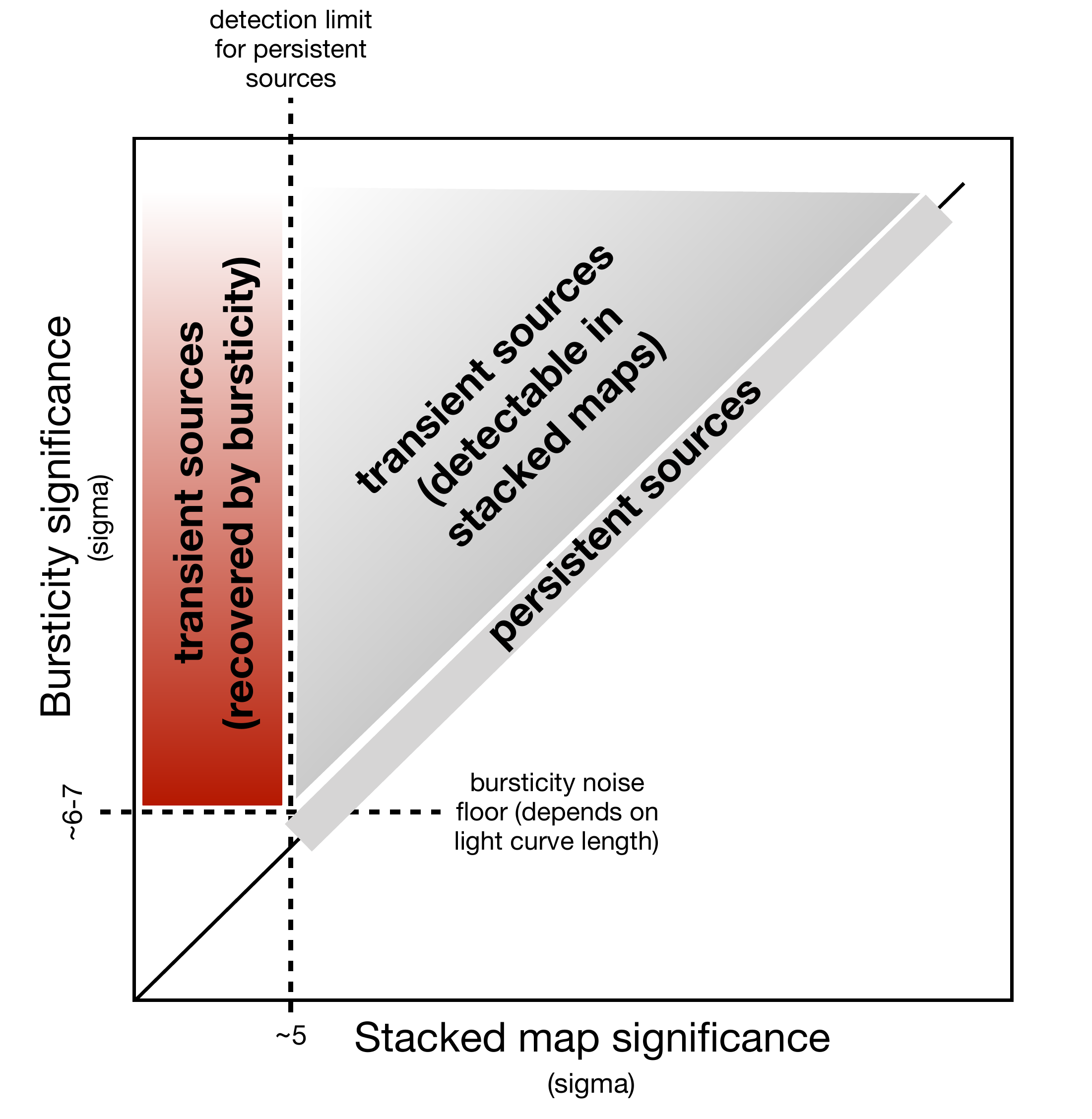}}
\caption{\footnotesize Source recovery by bursticity (explanation in the text).}
\label{fig:recovery}
\end{figure}

Fig.~\ref{fig:recovery} illustrates how the bursticity method aids the recovery of transient sources in long datasets, plotting the outburst significance against the significance derived from the full light curve. Persistent sources fall along the line of $y=x$ (the diagonal dashed white line), but many sources sit above $y=x$ indicating that their significance can be enhanced in a more limited time period. Sources that fall below the global significance threshold (the vertical red line) but above a burst detection threshold (horizontal dashed white line) can be recovered into the catalogue. The level of the burst detection threshold is determined experimentally - see below.

Bursticity searches, as currently implemented, are still somewhat biased because in order to improve the speed of the algorithm not all window sizes are tested and the stride (the speed with which the window passes along the light curve) is quite large. This means that not all possible windows are tested, although the assumption is made that the significance is only slowly degraded by using a non-optimal window, and only the very faintest outbursts will be missed in this way.

Of more concern is that bursticity is testing a very large number of non-independent windows, since the stride is typically $\sim10\%$ of the window length. This makes an analytical determination of the false alarm probability, or the burst detection threshold, difficult. Furthermore, the burst detection threshold depends both on the length and the time structure of the light curve. For the longest light curves (IBIS sources in the Galactic Plane) more than 100,000 window tests are performed during a search. In practice, monte-carlo simulations of a flux-randomised light curve with realistic temporal structure are used to establish confidence limits. Such tests are really only valid for light curves containing pure white noise, so any long-term source variability affects the determination of the burst detection threshold. Furthermore, the detection threshold actually increases with greater exposure as the number of trials increases, which is counter to normal expectations for persistent sources. 

\section{Follow-up campaigns of INTEGRAL surveys}

Multiwavelength followup of serendipitously detected X-ray sources is crucial to understand the properties of the objects observed, resulting in large imaging campaigns from radio frequencies to gamma-rays for specific areas of the sky. INTEGRAL provides input for many follow-up X-ray and optical campaigns. 

\subsection{Soft X-ray campaigns}

While the \integral\ surveys have been very successful at finding sources of high-energy emission, for new or previously poorly studied IGR sources, follow-up observations are necessary to obtain classifications.  \integral's few arcminute positions typically do not allow for the identification of optical or near-IR counterparts, especially in the crowded Galactic Plane, but follow-up X-ray observations reduce the error circles to the subarcsecond level (with the {\em Chandra X-ray Observatory}) or the few arcsecond level (with {\em XMM-Newton} or the {\em Neil Gehrels Swift Observatory}), allowing for the correct counterpart to be found.  In addition to localizations, soft X-ray spectra provide important diagnostics for classifying sources, including the spectral slope in the 1-10\,keV bandpass and the column density.  Finally, for some IGR sources, the angular resolution in the soft X-rays has led to the discovery of extended emission.

Early in the \integral\ mission, it was realized that many IGR sources were being found in the spiral arm regions of the Galaxy, and this led to the discovery of many HMXBs.  The first IGR source, IGR~J16318--4848, constitutes an excellent example where follow-up {\em XMM} observations provided important information.  While the {\em XMM} spectrum showed an extremely high column density $N_{\rm H}$$>$$10^{24}$\,cm$^{-2}$, the localization of the source indicated a match with a bright near-IR ($K_{s} = 7$) star, which proved that a large amount of material was obscuring the X-ray source \citep{mg03,walter03}. Spectroscopy of the near-IR source showed that it is a B[e]-type supergiant \citep{fc04}, surrounded by an inner cavity and an outer very large disc of gas and dust heated by this hot star, similar to Herbig Ae/Be stars, the compact object likely orbiting close to the rim separating the cavity from the disk \citep{2012ApJ...751..150C}. The existence of this disk was confirmed by the detection of flat-topped iron lines originating from a spherically symmetric disk wind, using broad-band spectroscopy with the ESO/VLT X-Shooter instrument \citep{2020ApJ...894...86F}. A stellar atmosphere and wind modeling, with the PoWR code, of the optical to mid-IR spectral energy distribution of this source -- adding mid-infrared {\it Spitzer} and {\it Herschel} data to these X-Shooter observations--, showed that the central star likely has an helium-enhanced atmosphere, due to an intense stellar wind shedding part of its hydrogen envelope \citep{2020ApJ...894...86F}.

A new class of obscured HMXBs had thus been discovered.  IGR~J16318--4848, with a likely long orbital period of $\sim 80$\,days \citep{iyer:2017}, is in the Norma spiral arm region of the Galaxy, and further soft X-ray observations uncovered more HMXBs in this region.  Although we still do not know the nature of the compact object in IGR~J16318--4848, IGR~J16320--4751 was found to be an HMXB with a slowly rotating (1300 sec period) neutron star using {\em XMM} observations \citep{lutovinov05, 2018A&A...618A..61G}.  Additional IGR HMXBs were uncovered in the Norma region as well as other part of the Galaxy, using {\em Chandra} localizations and information about the optical or near-IR counterpart \citep{tomsick06,tomsick08,tomsick09,tomsick12a,tomsick16,2020ApJ...889...53T}. Other identifications made use of the {\em Neil Gehrels Swift Observatory} \citep[e.g.][]{rodriguez:2008,2009A&A...494..417R,2009A&A...508..889R}, and {\em XMM-Newton} observations provided spectral and timing information about a large number of IGR HMXBs \citep{walter06,2014ApJ...795L..27H}.

In addition to HMXBs, other large groups of IGR sources are Cataclysmic Variables (CVs) and Low-Mass X-ray Binaries \citep[LMXBs, see e.g.][]{2018A&A...618A.150F,2020NewAR..9101547L}, Active Galactic Nuclei \citep[AGN, see e.g.][]{tomsick:2015}, and pulsars or Pulsar Wind Nebulae (PWNe).  In some cases, localizations by {\em Swift} provide identifications \citep[see e.g.,][and references therein]{2017MNRAS.470.1107L}.  CVs are often nearby with bright optical counterparts \citep{landi09a}, and AGN usually have IR or radio counterparts \citep{landi09b}.  For PWNe, extended X-ray emission is present, and IGR~J11014--6103 provides a dramatic example \citep{tomsick12b}.

While soft X-rays have been a critical component to classifying IGR sources, firm classifications most often require optical or near-IR spectroscopy, which is discussed in the following.


\subsection{Optical and near-infrared spectroscopy of IGR sources}
\label{sec:optic}

With the publication of the 1$^{\rm st}$ {\it INTEGRAL}/IBIS survey \citep{2004ApJ...607L..33B} it was realized that about one third of the catalogued hard X-ray sources had no identified nature or had too poor information on their characteristics. This percentage of unidentified or poorly known objects kept nearly constant in the subsequent issues of the all-sky {\it INTEGRAL}/IBIS surveys \citep[see][for details]{2010ApJS..186....1B}. Therefore, the need for a multiwavelength approach to pinpoint the nature of these objects, and ultimately the spectroscopic study of their optical and/or near-infrared (NIR) counterparts was apparent. Indeed, this technique allows the identification of the nature of the newly-discovered {\it INTEGRAL} sources and the characterization thereof by exploring their spectral features (mainly emission lines, which generally herald high-energy activity in cosmic sources) and overall continuum, thus permitting the determination of the main physical parameters for these sources, such as distance, luminosity and chemical composition among others.

However, the first, straightforward attempts to unravel the nature of these emitters involved searching the online multiwavelength repositories, such as SIMBAD\footnote{{\tt http://simbad.u-strasbg.fr}}, for known conspicuous (i.e., line-emitting) optical objects within the IBIS error circle of unidentified {\it INTEGRAL} sources: a paradigmatic example of the application of this technique was the case of IGR~J12349--6433 (=RT~Cru), for which a symbiotic star nature was first suggested by \cite{2005ATel..528....1M} on the basis of the localization of this peculiar optical source inside the hard X--ray positional uncertainty and of its published optical characteristics \citep{1994A&AS..106..243C}: the identification was then confirmed with the detection of soft X-rays \citep{2005ATel..591....1T} from this optical object.

\begin{figure*}
\centering
\begin{minipage}{0.8\textwidth}
\vspace{-0.45cm}
\includegraphics[trim= 0mm 0cm 0mm 0cm,
  width=1\textwidth,clip=t,angle=0.,scale=1.0]{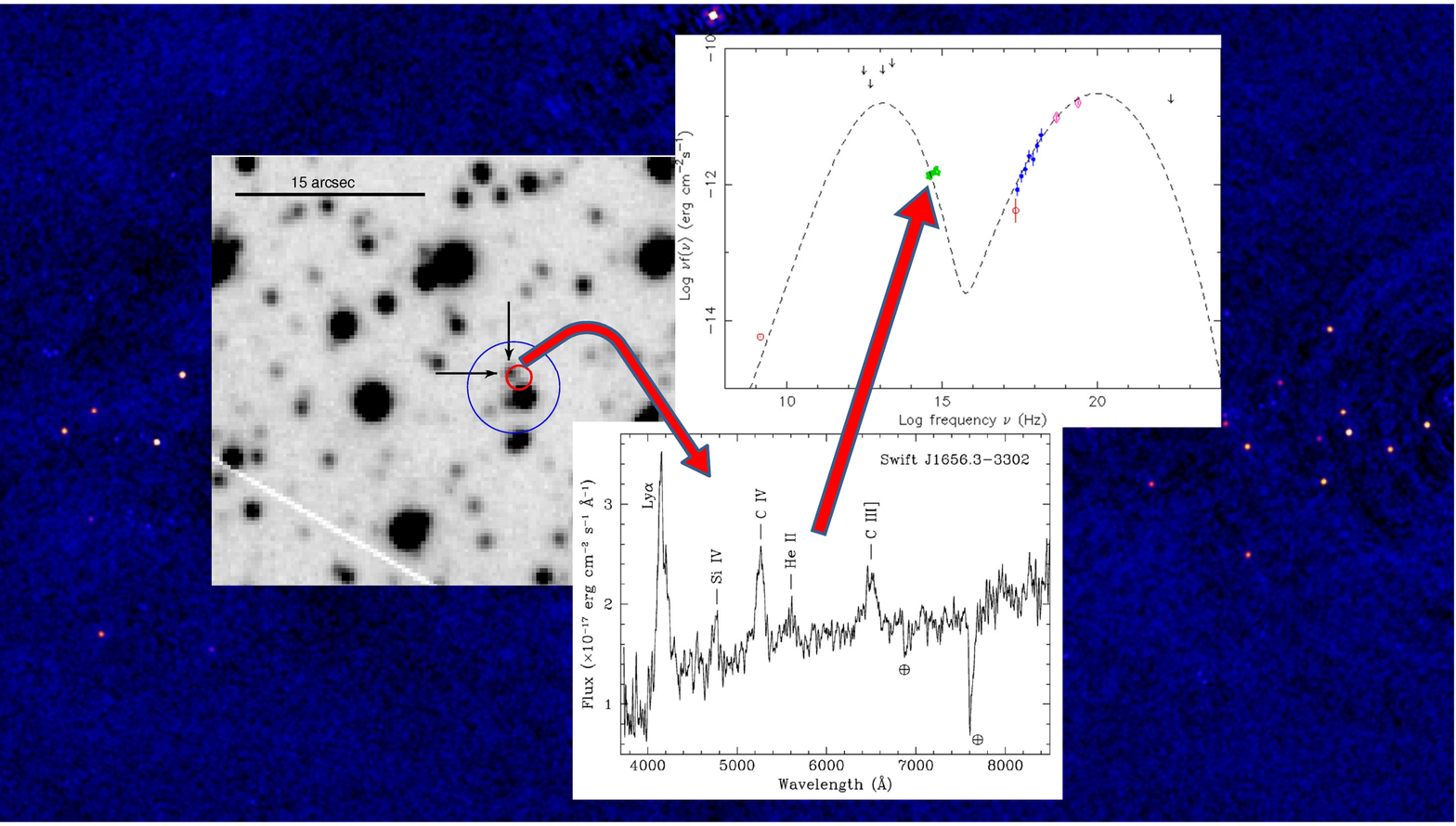}
\end{minipage}
\vspace{-1.0cm}
\caption{
Graphic representation of the followup process described in Sect.~\ref{sec:optic}, going from the accurate positioning through soft X--ray observations, to the optical spectroscopy
allowing the identification of the nature and main features of the source, and ending with
its multiwavelength characterization thanks to the building of a spectral energy 
distribution for the object \citep[adapted from][]{2008A&A...480..715M}.
\label{fig:followup}
}
\end{figure*}

Notwithstanding, the X-ray follow-up approach outlined in the previous subsection, of course, allows a better knowledge of the position of hard X-ray sources by reducing their error circles from a few arcminutes down to some arcseconds or less, thus reducing the search area in the sky by a factor of up to 10$^5$. Indeed, \cite{2006A&A...445..869S}  showed a low ($<$2\%) chance coincidence probability between the positions of {\it INTEGRAL} detections and those of softer X-ray sources within the hard X-ray error circle; similar figures are found using radio surveys \citep[e.g.,][]{2011MNRAS.416..531M}. This approach largely helps pinpointing the actual optical, as well as NIR, counterpart of the object responsible for the hard X--ray emission detected with {\it INTEGRAL}, which can then be studied through optical/NIR spectroscopy  \citep[see Fig.~\ref{fig:followup} for a sketch; for details, see][and references therein]{2008A&A...484..783C,zurita-heras:2008,butler:2009, 2013A&A...560A.108C,2018A&A...618A.150F,2013A&A...556A.120M,2006AstL...32..588B,2008AstL...34..653B,2008AstL...34..367B,2009AstL...35...71B,2012AstL...38....1L,2013AstL...39..513L,2012ApJ...761....4O,2018AstL...44..522K,2020AstL...45..836K}, and also through mid-IR observations \citep{rahoui:2008}.

\begin{figure*}
\centering
\begin{minipage}{0.75\textwidth}
\includegraphics[trim= 0mm 0cm 0mm 0cm,
  width=1\textwidth,clip=t,angle=0.,scale=0.98]{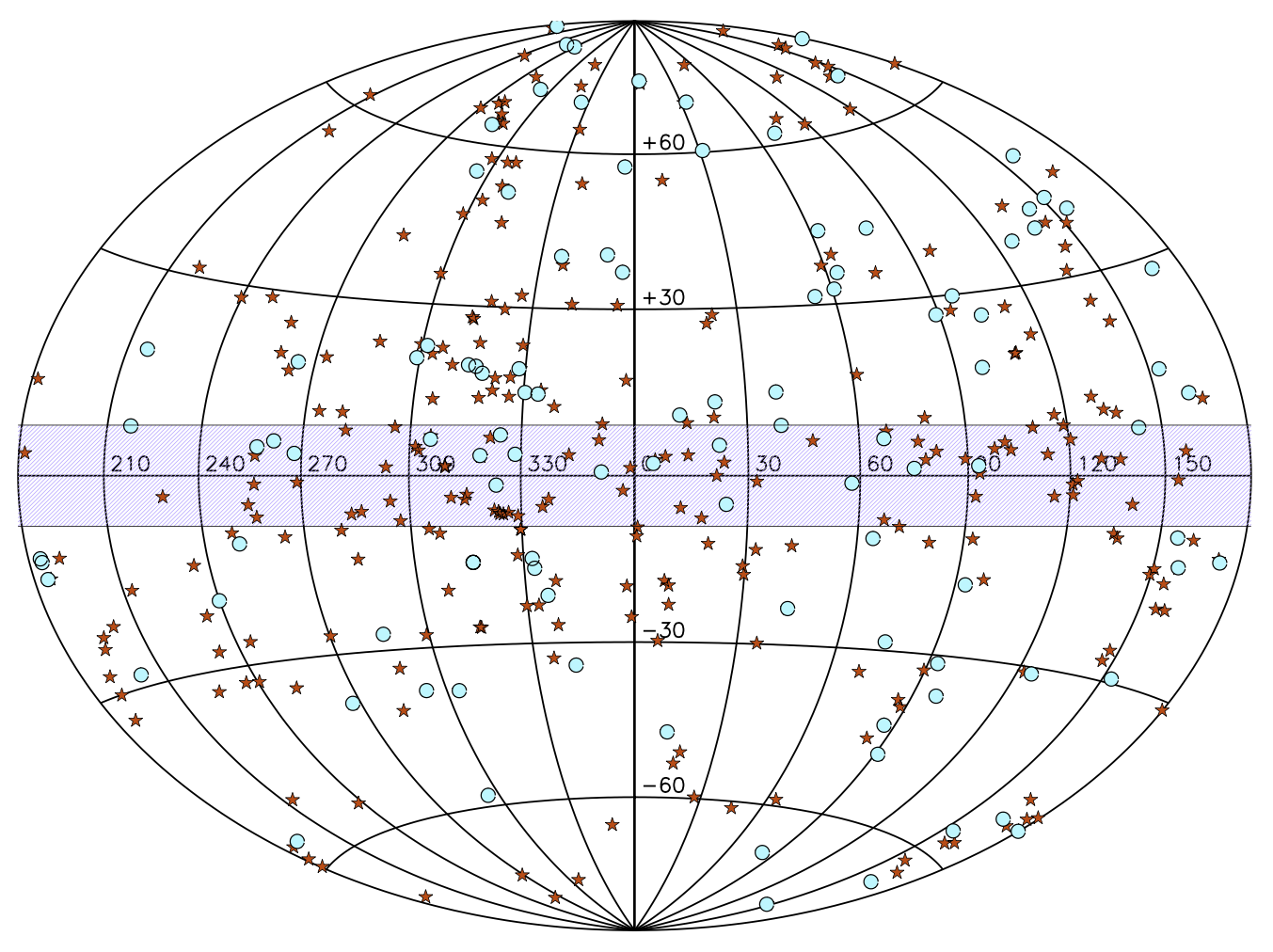}
\end{minipage}
\caption{AGNs detected by INTEGRAL/IBIS surveys \citep{2016MNRAS.460...19M}. The stars represent the 107 new active galaxies studied in \cite{2016MNRAS.460...19M} and first reported in the INTEGRAL/IBIS survey \citep{2016ApJS..223...15B}. The circles show AGNs detected in previous INTEGRAL/IBIS surveys. Adapted from \cite{2016MNRAS.460...19M}.
\label{fig:malizia2016}
}
\end{figure*}

Fifteen years of optical and NIR spectroscopic follow-up studies of unidentified
{\it INTEGRAL} sources performed by several groups worldwide led to a host of 
identifications: to the best of our knowledge, 265 such objects had their nature
identified or better described through optical/NIR spectroscopy, with the following
percentage breakdown: 58\% AGNs, 28\% Galactic X--ray binaries (3/4 of them identified 
as HMXBs), 13\% CVs and about 1\% active stars. We notice that,
if one takes into account the NIR spectroscopic identifications only, the overwhelming 
majority of sources is made of HMXBs ($\sim$90\%), with just about 10\% of AGNs
\citep{2013A&A...560A.108C, 2018A&A...618A.150F} and references therein. A large fraction of the INTEGRAL AGN have been characterized in terms of optical and X-ray properties by \cite{2016MNRAS.460...19M}, as demonstrated in the all-sky map in Fig.~\ref{fig:malizia2016}.

The above figures can be compared to e.g. the identified {\it INTEGRAL} sources in the first IBIS catalogue \cite{2004ApJ...607L..33B} grouped into the same broad classes: this gives AGN, X-ray binary and CV percentages of 4\%, 64\% (with LMXBs being more than twice in number than HMXBs) and 3\% and no active stars, respectively. Thus, the contribution of {\it INTEGRAL} to the advancement of our knowledge of the hard X--ray sky, combined with the optical/NIR followup of the unidentified sources in its surveys, has been multifold, namely: (1) it allowed the exploration of the extragalactic sky through the so-called 'Zone of Avoidance' along the Galactic Plane, where the attenuation induced by dust and gas is not an hindrance for high-energy detectors; (2) similarly, it enhanced our knowledge on HMXBs by increasing the source statistics; (3) it allowed the detection of a siginficant number of (mostly magnetic) hard X--ray emitting CVs, which was unexpected \citep[but not unprecedented in hindsight, see, e.g., recent review of][and referenes therein]{2020NewAR..9101547L}.

Although in-depth presentations will be given in other contributions within this group of reviews, we would like to conclude this section by focusing on a few issues raised thanks to the optical/NIR follow-up of {\it INTEGRAL} sources: these considerations are connected with the points listed above and, according to us, deserve to be mentioned.

First, it is stressed \citep{2012A&A...538A.123M} that the use of medium-sized and large telescopes (above 4 metres in diameter) allows the study of the faint end of the distribution of putative optical counterparts of the extragalactic share of these high-energy sources. Indeed, a Kolmogorov-Smirnov test showed that the probability that the redshift distributions of the newly-identified hard X--ray AGNs and of the ones already classified in the {\it INTEGRAL} surveys are the same is less than 0.001; that is, the former ones are are drawn from a different distribution of more distant objects. Therefore, the deeper {\it INTEGRAL} observations available with the latest surveys allow one to explore the hard X-ray emitting sources in the far universe, at an average distance $\sim$5 times larger than that of the average of such type of objects known up to now.

Then, we remark that this identification program in hard/soft X--rays plus optical suprisingly boosted the number of magnetic CVs, suggesting this subclass as an important member of Galactic X--ray sources: these systematic studies, according to the review of \cite{2020AdSpR..66.1209D}, allowed increasing the sample the magnetic CV subclass by a factor of two, permitting extensive and dedicated explorations on specific cases, on these objects as a group, as well as comparative studies as a function of the magnetic field strength of the white dwarf accretor.

Finally, we briefly mention that the combination of the information extracted from the {\it INTEGRAL} surveys and the optical follow-up work allowed the discovery and/or the characterization of a number of new classes of Galactic X-ray binaries, such as Supergiant Fast X-ray Transients \citep{2005A&A...444..221S,2006ESASP.604..165N}, Transitional Millisecond X--ray Pulsars \citep[e.g.,][]{2013A&A...550A..89D,2014MNRAS.441.1825B} and Symbiotic X-ray Binaries \citep[e.g.,][]{2007A&A...470..331M,2012MNRAS.422.2661S}.

The timely multiwavelength exploration of the {\it INTEGRAL} sources remaining to be identified and classified is thus of high importance.


\section{Total CXB spectrum measurements by \integral\ in the 4-200 keV band}

\integral/IBIS \citep{ibis} together with Swift/BAT \citep{2005SSRv..120..143B}, having both good sensitivity and wide-field sky coverage,  allowed to make  a significant progress in the study of the high energy domain in the last decay. In particular they have provided a great improvement in our knowledge of the extragalactic sky by detecting more than 1000 (mostly local) AGN at high energies. In Fig.~\ref{fig:malizia2016} all the AGN detected by \integral/IBIS until 2016, and consequently classified and spectrally characterised, have been plotted \citep{2016MNRAS.460...19M}. The high energy band (20--200~keV) is extremely important for the study of the extragalactic sky and in particular for AGN; it is also the most appropriate for population and survey studies, since it is almost unbiased against obscuration, a severe bias which affects surveys at other frequencies. The great improvement achieved in this field thanks to the \integral\ surveys has been extensively discussed in the dedicated review on INTEGRAL view of AGN  by Malizia et al. in this special issue. Here we want to stress the importance of the survey work for the determination of the Cosmic X-Ray Background (CXB).

As described above, in the deepest extra-galactic fields \integral\ can reliably detect sources at the limiting 20--60~keV flux $\sim {\rm few~} 10^{-12} ~{\rm erg\,s^{-1}\,cm^{-2}}$ \citep[e.g.][]{2016MNRAS.459..140M}. At this sensitivity level only a small fraction ($\sim\,$few~\%) of the total CXB in this energy band is resolved \citep{krivonos2007a}. Detection of fainter objects comprising the bulk of the CXB requires a prohibitively long exposure time. This stems from the intrinsic limitation of coded mask telescopes, when photons from a single source are distributed across the entire detector. At the same time, coded mask telescopes often have large FOVs and the total count rate associated with the CXB can be large. Therefore, the detector spectra of the telescopes on-board INTEGRAL contain information on the total flux from all resolved and unresolved (no matter how faint they are) sources within the FOV. 

Separating the contribution of the CXB from the particle background in the detector spectra is a non-trivial exercise. One (difficult) route is to build a comprehensive model of the detector non-astrophysical background, which usually dominates the count rate, so that this model can be subtracted from recorded spectra. Another possibility is to modulate the astrophysical flux so that it can be singled out from the particle background. Here we discuss the results of total  CXB flux measurements by \integral\ using the latter possibility.

The modulation of the CXB flux has already been employed in early space X-ray experiments. In particular the HEAO-1 observatory used a movable 5~cm thick CsI crystal to partly block the instrument field of view and to modulate the CXB signal \citep{1997ApJ...475..361K,1999ApJ...520..124G}. For \integral, the problem of modulation was solved by "placing" the Earth into the telescopes FOVs \citep{2007A&A...467..529C,2010A&A...512A..49T}. The same technique was also used by \citet{2007ApJ...666...86F}  for the BeppoSAX mission and by \citet{2008ApJ...689..666A} for the data of the Burst Alert Telescope (BAT) aboard the Swift spacecraft.

\begin{figure}
\includegraphics[trim= -10mm 0cm -10mm 0cm,
  width=1\textwidth,clip=t,angle=0.,scale=0.5]{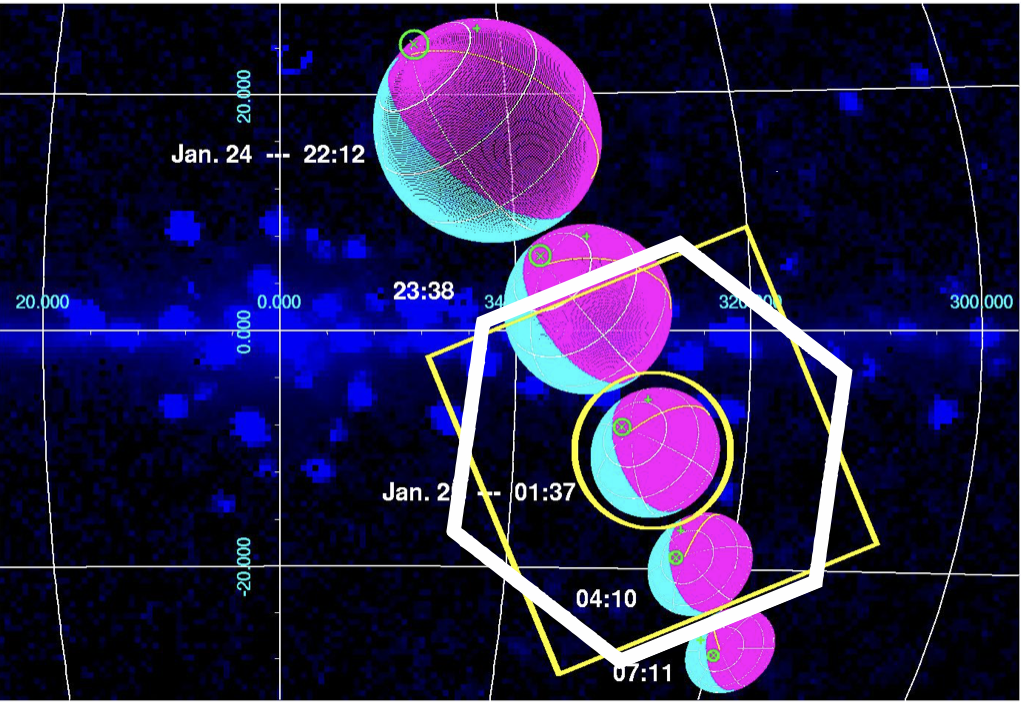} 
\caption{Illustration of the \integral\ Earth observing mode in 2006 \citep[see][for the original version of this figure]{2007A&A...467..529C}. FOVs of JEM-X, IBIS and SPI are schematically shown with a circle, a box and an hexagon respectively superposed on to the RXTE 3-20 keV slew map in Galactic coordinates.  In the course of this observation the pointing direction of the telescopes remains the same, while the Earth moves across the instruments FOVs. The day side of the Earth is shown by cyan color. As the distance from the Earth increases during this portion of the 3-day INTEGRAL orbit, the angular size of the Earth disk decreases and the the solid angle of the obscured CXB goes down.
\label{fig:earth_track}
}
\end{figure}

\integral\ was launched onto a 3-day elongated orbit with a perigee of $\sim\!6\,000\,{\rm km}$ and an apogee of $\sim 150\,000\,{\rm km}$. The orbit crosses the Earth radiation belts at the distance of $\sim 50\,000\,{\rm km}$, so that useful observations are possible at larger distances. During \integral\ observations of the Earth in 2006, the satellite was kept in a controlled 3-axis stabilization with telescopes' axes staring at a point where the Earth was predicted to be some 6 hours after the perigee passage (see Fig.~\ref{fig:earth_track}). Four such observations were performed in 2006, each lasting about 30~ks ($\sim$ 8 hours). When the Earth was close to the telescopes' pointing direction, the radius of the Earth disk was $\sim 5.4^\circ$, i.e. the CXB signal from some 90~sq.deg. was subtended. In terms of the flux near 30 keV, such solid angle corresponds to $\sim200\,{\rm mCrab}$, i.e. a very significant modulation. Moreover, as the distance from the Earth changes and the Earth moves through the FOV, the modulation amplitude can be readily predicted and, therefore, used to separate it from other contaminating signals.

Apart from the genuine CXB signal, modulated by the obscuration by the Earth disk, there is a number of other variable components that have to be accounted for. Namely:
\begin{itemize}
    \item Individual compact X-ray sources (mostly in the Galaxy) which induce sharp edges in the recorder light-curves when the disk edge goes over them. 
    \item Unresolved foreground emission of the Galactic Ridge.
    \item Emission of the Earth atmosphere due to scattering of the CXB photons \citep{2008MNRAS.385..719C} and induced by cosmic rays impinging the atmosphere \citep{2007MNRAS.377.1726S}.
    \item Earth Auroral emission that was strong and variable in several \integral\ observations.
    \item Variability of the intrinsic detector background.
\end{itemize}

\begin{figure}
\begin{minipage}{0.49\textwidth}
\includegraphics[trim= 0mm 1cm 0mm 0cm,
  width=1\textwidth,clip=t,angle=0.,scale=0.98]{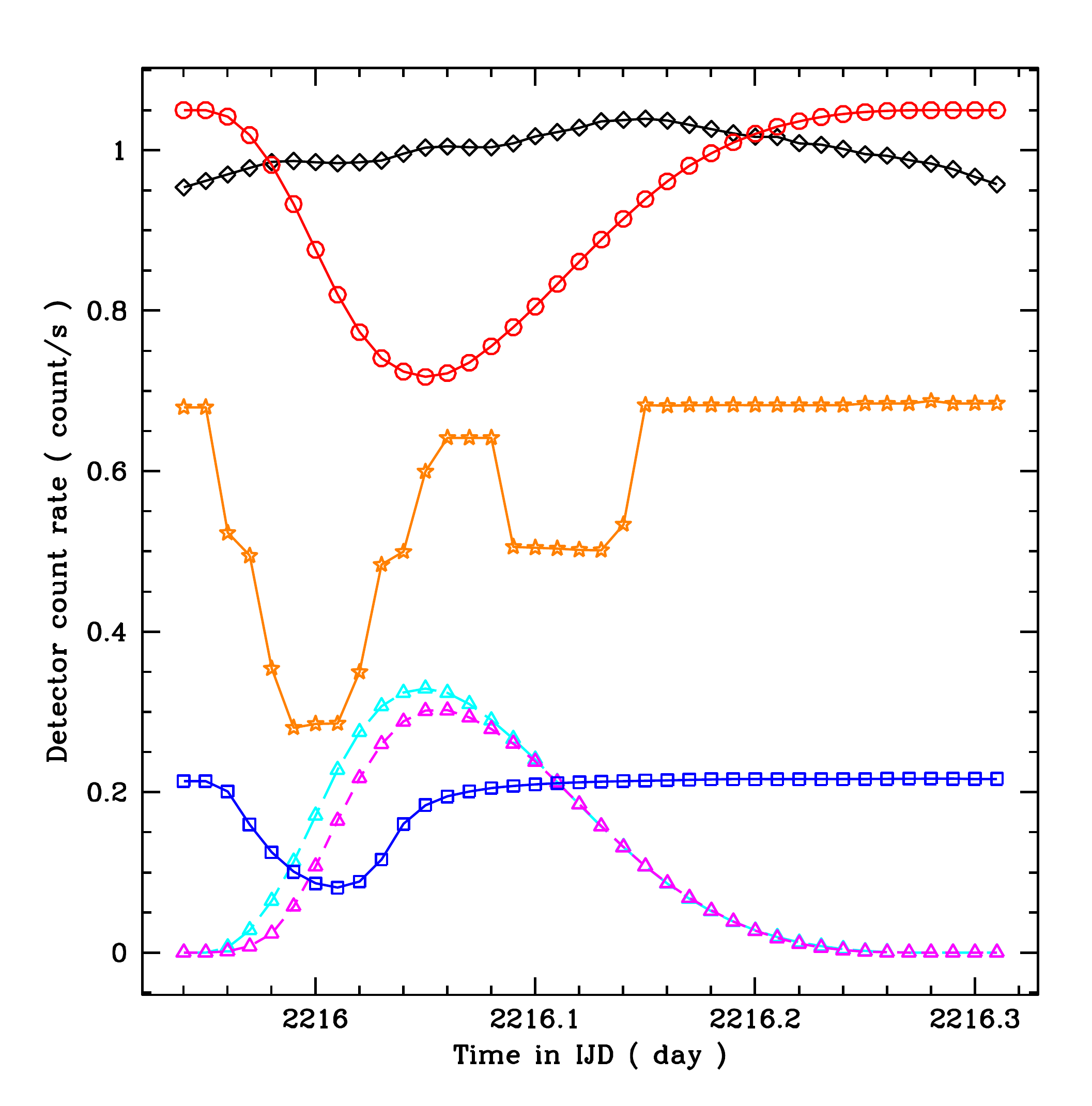}
\end{minipage}
\caption{Model lightcurves of each component for a $\sim 2$~keV wide energy channel centered at $\sim 27$~keV during the first \integral\ observation of the Earth. The drop of the detector count rate due to the shadowing of the CXB is shown with red circles.  The other components are the effective contribution of the point sources (orange stars), the GRXE (blue squares), the Earth CXB reflection (cyan triangles, long-dashed), the CR-induced emission (magenta triangles, short-dashed), and the estimated instrumental background (black diamonds). Adapted from \citet{2010A&A...512A..49T}.
\label{fig:earth_lc_model}
}
\end{figure}

Individual compact X-ray sources are relatively easy to deal with, as long as they are not variable. Their spectra can be measured using a portion of the observation when they are not obscured by the Earth disk. Their modulation pattern can be described as a simple mask function (either 0 or 1 at any moment) set by their position with respect to the Earth disk (see Fig.~\ref{fig:earth_lc_model}).

The Galactic Ridge emission is more difficult to account properly due to its diffuse nature. The portions of the data when the Earth disk was moving over the bright regions of the Ridge can be either ignored \citep{2007A&A...467..529C}, or a spatial model of the Ridge can be used \citep{2010A&A...512A..49T}.

\begin{figure}
\begin{minipage}{0.49\textwidth}
\includegraphics[trim= 0mm 0cm 0mm 0cm,
  width=1\textwidth,clip=t,angle=0.,scale=0.98]{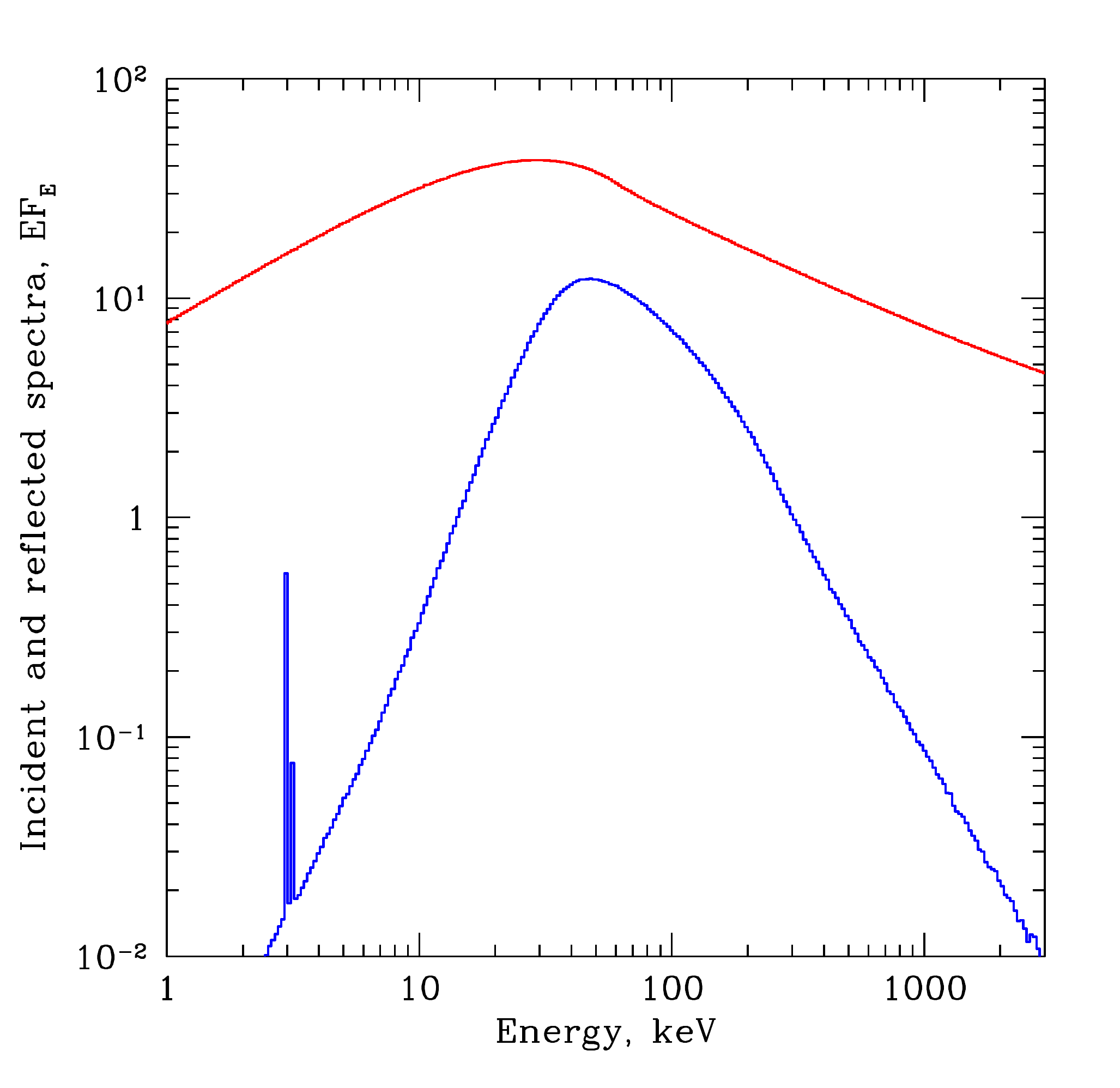}
\end{minipage}
\caption{The CXB spectrum (top) and the spectrum reflected by the Earth atmosphere (bottom). The reflected spectrum was integrated over all angles. The features in the reflected spectrum near 3 keV are the fluorescent lines of Argon. Adapted from \citet{2008MNRAS.385..719C}.
\label{fig:cxb_albedo}
}
\end{figure}

The idea of having the Earth shadowing the CXB is based on the assumption that its disk is dark in X-rays. This is only partly true, since the CXB photons can be partly reflected by the Earth atmosphere, while cosmic rays can generate secondary gamma-radiation. The CXB radiation reflected by the atmosphere (Fig.~\ref{fig:cxb_albedo}) can be straightforwardly calculated \citep{2008MNRAS.385..719C}, although it depends on the CXB spectrum itself. Fortunately, the albedo, i.e. the ratio of the reflected and the incident spectra is weakly sensitive to the shape of the incident spectrum in the relevant energy band, implying that the reflected component can be readily evaluated if the albedo is calculated using reasonable guess on the CXB shape. 

\begin{figure}
\begin{minipage}{0.49\textwidth}
\includegraphics[trim= 0mm 0cm 0mm 0cm,
  width=1\textwidth,clip=t,angle=0.,scale=0.98]{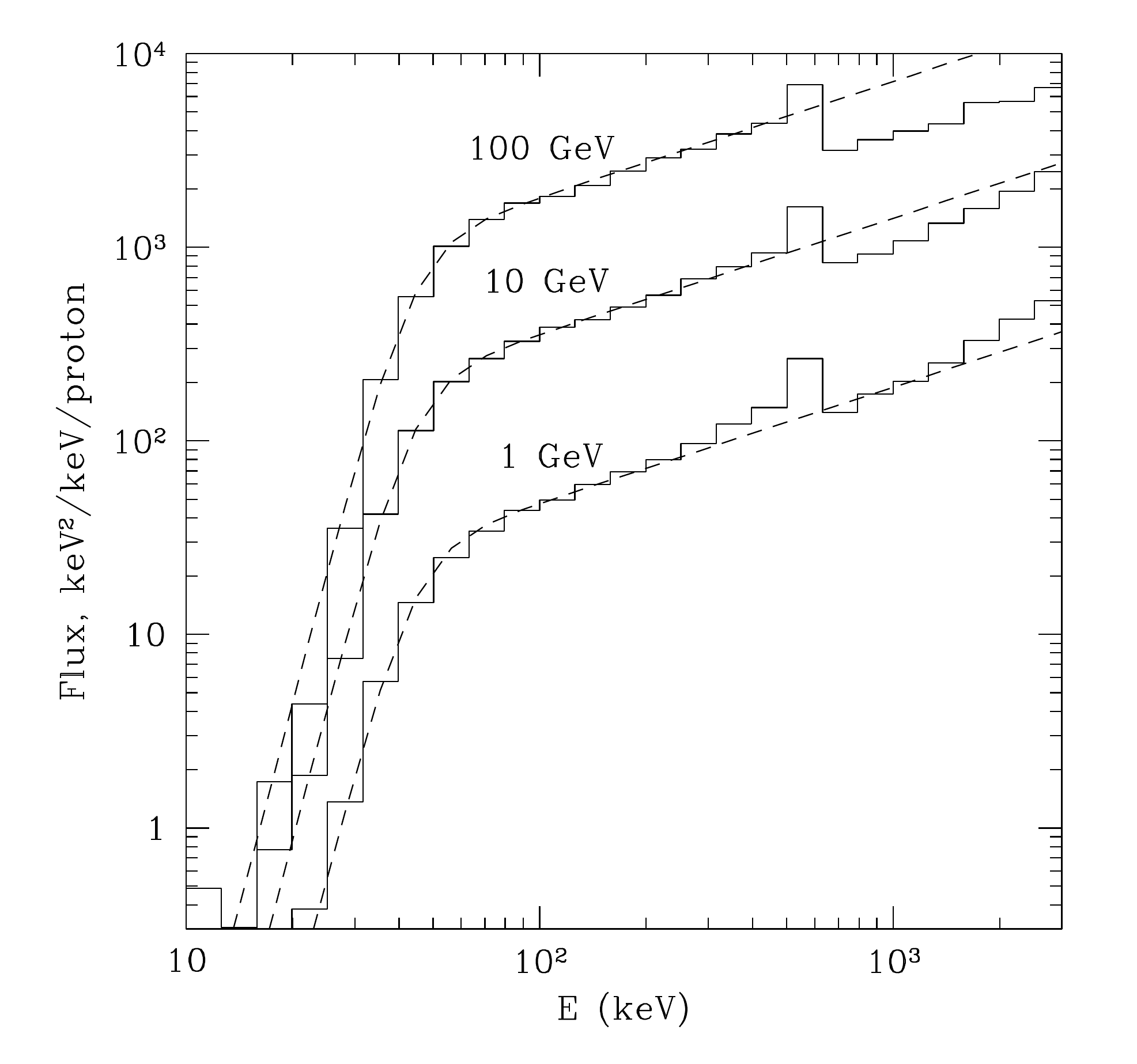}
\end{minipage}
\caption{Examples of simulated spectra (solid lines) of atmospheric emission produced by cosmic protons of given energy: $E_p=$ 1, 10 and 100 GeV. It can be seen that in the photon energy range 25-300 keV the shape of the emergent spectrum is almost invariant (the dashed lines). Adapted from \citet{2007MNRAS.377.1726S}.
\label{fig:cxb_crs}
}
\end{figure}

Cosmic rays, in particular protons impinging the Earth atmosphere, undergo a series of hadronic interactions and electromagnetic cascades to induce a glow of the atmosphere in hard X-rays. Examples of expected spectra from the Monte Carlo simulations \citep{2007MNRAS.377.1726S} based on the GEANT4 software package \citep{geant4} are shown in Fig.~\ref{fig:cxb_crs}.

Finally, the emission of the Earth Aurora can be bright and highly variable. Currently, there are no good recipes for properly modeling the contribution of the Aurora. Therefore all observations severely affected by the Aurora were excluded from the analysis. For instance, among four 30~ks \integral\ observations of the Earth done in 2006, two have signatures of the Aurora emission.

With all the components mentioned above, the spectrum observed by the \integral\ instruments $S(E,T)$ at any given moment can be represented as
\begin{eqnarray}
S(E,t)=B(E,t)+ \nonumber \\
CXB(E)-CXB(E)\times\Omega(t)\times [1-A(E)]+ \nonumber \\
CR(E)\times \Omega(t)+\nonumber \\
\Sigma_i I_i(E,\alpha,\delta)\times M(\alpha,\delta,t),
\label{eq:earth_lc}
\end{eqnarray}
where $B(E,t)$ is the intrinsic background; $CXB(E)$ is the spectrum of the CXB, $\Omega(t)$ is the solid angle subtended by the Earth disk, $A$ is the Earth albedo for the CXB-like spectrum (Fig.~\ref{fig:cxb_albedo}), $CR(E)$ is the particle-induced emission of the Earth atmosphere (Fig.~\ref{fig:cxb_crs}), $I_i(E,\alpha,\delta)$ is the spectrum of individual bright source located at coordinates $(\alpha,\delta)$ and $M(\alpha,\delta,t)$ is the time dependent mask due to the Earth occultation of the source position. 

The examples of the light curves predicted for the ISGRI/IBIS instrument in a narrow energy band near 27 keV are shown in Fig.~\ref{fig:earth_lc_model}. As is clear from this Figure and also from eq.~(\ref{eq:earth_lc}), the CXB obscuration, the CXB and CR albedos share the same time dependence associated with $\Omega(t)$. Therefore, these components have to be modeled simultaneously using the information about their spectral shapes (see Figs.~\ref{fig:cxb_albedo} and \ref{fig:cxb_crs}).

\begin{figure}
\begin{minipage}{0.49\textwidth}
\includegraphics[trim= 0mm 0cm 0mm 0cm,
  width=1\textwidth,clip=t,angle=0.,scale=0.98]{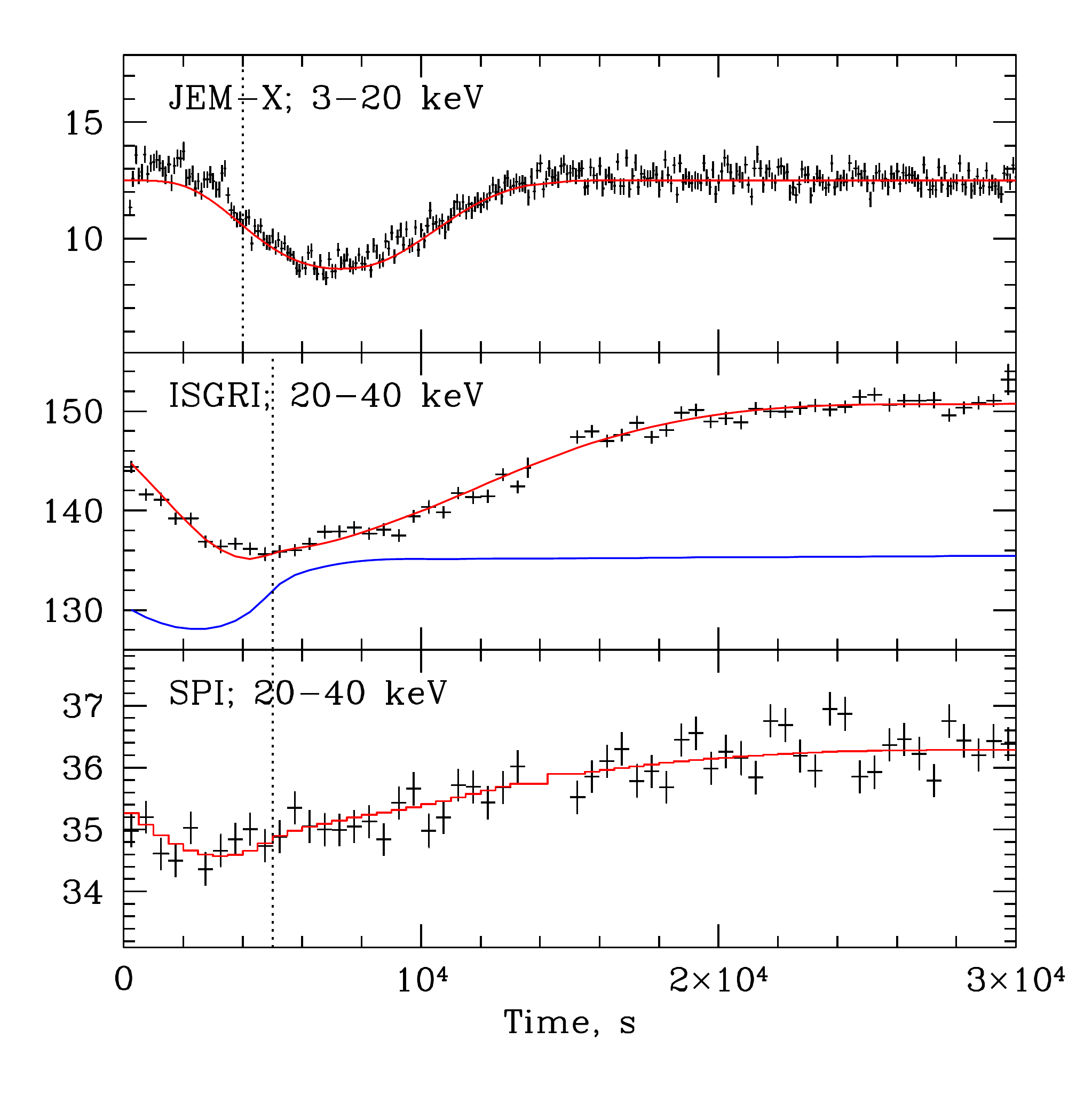}
\end{minipage}
\caption{The light curves (crosses) of JEM-X, IBIS/ISGRI and SPI instruments in units of
counts per second. The red curves show the model light curve that includes the shadowing by the Earth disk. In the middle panel the blue curve shows schematically (with arbitrary normalization) the time dependence of the Galactic Ridge emission, modulated by the Earth occultation. In order to avoid contamination of the CXB measurements due to Galactic plane contribution, the first few ksec of data (on the left of the dotted vertical lines) were dropped from the analysis. Note that for JEM-X a less strict cut was applied since its field of view is smaller than that of the other instruments. Adapted from \citet{2007A&A...467..529C}.
\label{fig:earth_lc_observed}
}
\end{figure}

Actual light curves for all three instruments on board the \integral\ are shown in Fig.~\ref{fig:earth_lc_observed}. It is clear that the detector light curves are indeed modulated and the time variations are consistent with expectations (red curves). Repeating the analysis in many different bands and measuring the amplitude of the modulation allows the reconstruction of the CXB spectrum \citep[see][for details of the analysis]{2007A&A...467..529C,2010A&A...512A..49T}.

\begin{figure}
\begin{minipage}{0.49\textwidth}
\includegraphics[trim= 0mm 0cm 0mm 0cm,
  width=1\textwidth,clip=t,angle=0.,scale=0.98]{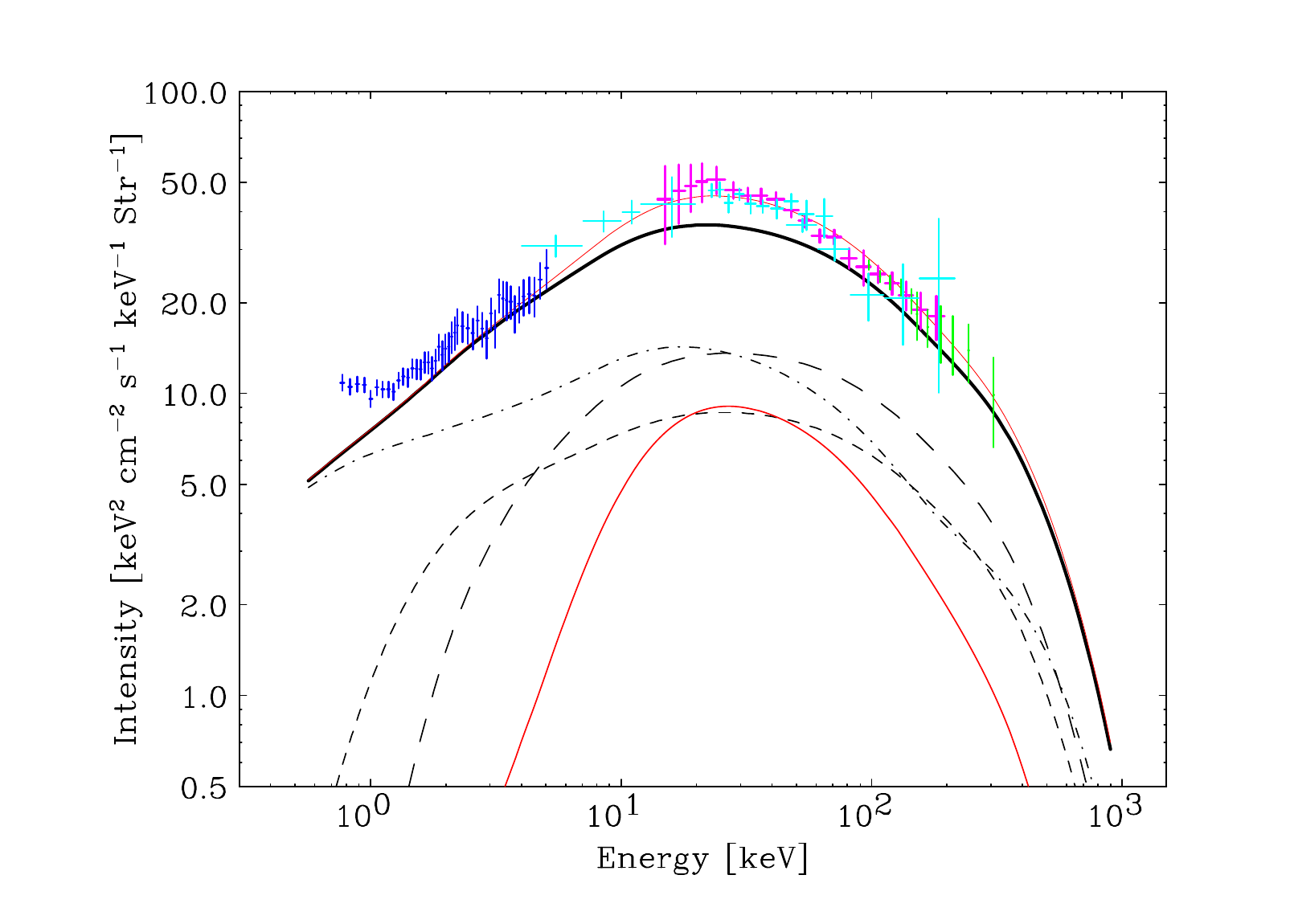}
\end{minipage}
\caption{Cosmic X-ray Background (CXB) spectrum calculated from AGN population synthesis models \citep[upper solid curve, red;][]{2014ApJ...786..104U} compared with the observed data by different X-ray missions \citep{2008ApJ...689..666A}. Middle solid curve (black): the integrated spectrum of Compton-thin AGNs (log \nh\ $<$24). Lower solid curve (red): that of Compton-thick AGNs (log \nh = 24--26). Long-dashed curve (black): that of AGNs with log \nh = 23--24. Short-dashed curve (black): that of AGNs with log \nh = 22--23. Dot-dashed curve (black): that of AGNs with log \nh $<$22. Data points in the 0.8--5 keV (blue), 4--215 keV (cyan), 14--195 keV (magenta), and 100--300 keV (green) bands refer to the CXB spectra observed with ASCA/SIS \citep{1995PASJ...47L...5G}, \integral\ \citep{2007A&A...467..529C}, Swift/BAT \citep{2008ApJ...689..666A}, and HEAO A4 \citep{1999ApJ...520..124G}, respectively. Adapted from \citet{2014ApJ...786..104U}.
\label{fig:cxb_earth}
}
\end{figure}

The derived CXB spectrum is shown in Fig.~\ref{fig:cxb_earth} along with the data from other experiments, including HEAO1 A4 in the 100--300~keV band \citep{1999ApJ...520..124G} and {\em Swift}/BAT in the 14--195~keV band \citep{2008ApJ...689..666A}. The obtained normalization of CXB spectrum is $\sim$10\% higher than suggested by \cite{1999ApJ...520..124G} and consistent with recent CXB measurement performed with the \nustar\ telescope in 3--20~keV band  \citep{2020arXiv201111469K}. The observed CXB spectrum is well described by the standard population synthesis model of AGNs, including the fraction of Compton-thick AGNs and the reflection strengths from the accretion disk and torus based on the luminosity- and redshift- dependent unified scheme \citep{2014ApJ...786..104U}. 

\section{Conclusions}

One of the many areas where the \integral\ observatory provides a significant scientific outcome to the astrophysical community is surveying the sky at energies above 20 keV. 
\integral\ surveys of the Galactic Plane and extragalactic fields triggered a large number of new studies and observational campaigns in other wavelengths. 

Thanks to its coded-aperture design, the IBIS telescope, the main instrument for \integral\  hard X-ray surveys, incorporates a very large fully-coded FOV of $28^{\circ}\times28^{\circ}$, which allows to conduct cartography of the sky in reasonable time. In particular, \integral\ is able to take hard X-ray snapshots of the whole Milky Way over a time scale of a year, which is far from the capabilities of narrow-FOV grazing X-ray telescopes.

Apart from providing the census of hard X-ray emitters over the whole sky, \integral\ conducted a unique observation of the large-scale cosmic X-ray background via Earth-occultation manoeuvre, which will undoubtedly be included in the legacy of the \integral\ observatory.



\section*{List of abbreviations}

\noindent  List of definitions of abbreviations used in the paper.\\
FOV: Field of View;\\
AGN: Active Galactic Nuclei;\\
HMXB: High Mass X-ray Binary;\\
LMXB: Low Mass X-ray Binary;\\
CV: Cataclysmic Variable;\\
PSR: Pulsar;\\
PWN: Pulsar Wind Nebula;\\
CXB: Cosmic X-Ray Background;\\
GRXE: Galactic Ridge X-ray Emission;\\
PoWR: the Potsdam Wolf-Rayet Models.

\section*{Acknowledgements}

We would like to thank all our colleagues who contributed over the years to INTEGRAL data analysis and the interpretation. This review is based on observations with INTEGRAL, an ESA project with instruments and the science data centre funded by ESA member states (especially the PI countries: Denmark, France, Germany, Italy, Switzerland, Spain), and Poland, and with the participation of Russia and the USA. RK, EC and RS acknowledge support from the Russian Science Foundation grant 19-12-00369 in working on this review. The Italian co-authors acknowledge support from the Italian Space Agency (ASI) via different agreements including the last ones, 2017-14-H.0 and 2019.HH-35-HH.0. JAT acknowledges partial support from NASA through {\em Chandra} Award Number GO8-19030X issued by the {\em Chandra} X-ray Observatory Center, which is operated by the Smithsonian Astrophysical Observatory.

\section*{References}
\bibliography{main}
\end{document}